\documentclass[useAMS,usenatbib]{mn2e}

\usepackage{color}
\usepackage{graphicx}
\usepackage[below]{placeins}
\usepackage{natbib}
\usepackage{journal_shortcuts}
\bibpunct{(}{)}{;}{a}{}{,}

\def\gtsim{\mathrel{\hbox{\rlap{\hbox{\lower4pt\hbox{$\sim$}}}\hbox{$>$}}}}
\def\lesssim{\mathrel{\hbox{\rlap{\hbox{\lower4pt\hbox{$\sim$}}}\hbox{$<$}}}}
\def\Msunpyr{M$_{\odot}\,$yr$^{-1}$}
\def\Msun{M$_{\odot}$}
\def\A{{\rm\thinspace \AA}}
\def\cm{{\rm\thinspace cm}}

\def\erg{{\rm\thinspace erg}}

\def\ga{{\rm\thinspace gauss}}

\def\km{{\rm\thinspace km}}
\def\kpc{{\rm\thinspace kpc}}

\def\Msun{\hbox{$\rm\thinspace M_{\odot}$}}

\def\s{{\rm\thinspace s}}
\def\ps{{\rm\thinspace s^{-1}}}

\def\yr{{\rm\thinspace yr}}
\def\ergpscmps{\hbox{$\erg\cm^{-2}\s^{-1}\,$}}

\def\ergpscmpspa{\hbox{$\erg\cm^{-2}\s^{-1}\A^{-1}\,$}}
\def\ergps{\hbox{$\erg\s^{-1}\,$}}

\def\kmps{\hbox{$\km\ps\,$}}

\def\Msunpyr{\hbox{$\Msun\yr^{-1}\,$}}

\def\pyr{\hbox{$\yr^{-1}\,$}}

\def\hb{\hbox{{\rm H}$\beta$}}

\def\oii{\hbox{{\rm [O{\sc ii}]}}}

\def\h0{\hbox{{\rm H}$^0$}}
\DeclareMathAlphabet{\vib}{OML}{cmm}{m}{it}

\title[The cool-core of RXCJ1504.1-0248]{Central galaxy growth and feedback in the  most massive nearby cool core cluster}
\author[G. A. Ogrean et al.]{G. A. Ogrean$^{1}$\thanks{E-mail: g.ogrean@jacobs-university.de}, N. A. Hatch$^{2,6}$, A. Simionescu$^{3}$, H. B\"{o}hringer$^{3}$, M. Br\"{u}ggen$^{1}$,\and A. C. Fabian$^4$, N. Werner$^{5}$\\
$^{1}$Jacobs University Bremen, Campus Ring 1, D-28759 Bremen, Germany\\
$^{2}$Leiden Observatory, Niels Bohrweg 2, 2333 CA, The Netherlands\\
$^{3}$Max-Planck-Institut f\"{u}r extraterrestrische Physik, Giessenbachstrasse, D-85741 Garching, Germany\\
$^{4}$Institute of Astronomy, Madingley Road, Cambridge CB3 0HA\\
$^{5}$KIPAC, Stanford University, 382 Via Pueblo Mall, Stanford, CA 94305, USA\\
$^{6}$School of Physics and Astronomy, The University of Nottingham, University Park, Nottingham NG7 2RD, UK \\}
\begin{document}

\pagerange{\pageref{firstpage}--\pageref{lastpage}} \pubyear{2002}

\maketitle

\label{firstpage}

\begin{abstract}
We present multi-wavelength observations of the centre of RXCJ1504.1-0248 -- the galaxy cluster with the most luminous and relatively nearby cool core at $z\sim 0.2$. Although there are several galaxies within 100\,kpc of the cluster core, only the brightest cluster galaxy (BCG), which lies at the peak of the X-ray emission, has blue colours and strong line-emission. Approximately 80\Msunpyr\  of intracluster gas is cooling below X-ray emitting temperatures, similar to the observed UV star formation rate of $\sim$140\Msunpyr. Most star formation occurs in the core of the BCG and in a 42\,kpc long filament of blue continuum, line emission, and X-ray emission, that extends southwest of the galaxy. 

The surrounding filamentary nebula is the most luminous around any observed BCG. The number of ionizing stars in the BCG is barely sufficient to ionize and heat the nebula, and the line ratios indicate an additional heat source is needed. This heat source can contribute to the H$\alpha$-deduced star formation rates (SFRs) in BCGs and therefore the derived SFRs should only be considered upper limits.

AGN feedback can slow down the cooling flow to the observed mass deposition rate if the black hole accretion rate is of the order of 0.5 \Msunpyr at 10\% energy output efficiency.

The average turbulent velocity of the nebula is $v_{\rm turb}\sim 325\,\kmps$ which, if shared by the hot gas, limits the ratio of turbulent to thermal energy of the intracluster medium to less than 6 percent. 

\end{abstract}


\section{Introduction \label{intro}}

Cool core clusters are characterised by  bright central peaks in their X-ray surface brightness profiles, cool and radially-increasing core temperatures, and short central cooling times. Unless the gas is reheated, the large amount of energy radiated in X-rays  implies several hundred solar masses of gas cools below $10^{7}$\,K every year \citep{CowieBinney77,FabianNulsen77}.  However, X-ray spectra from \emph{XMM-Newton} and \emph{Chandra} show cool-core clusters lack strong cooling lines, so less gas is cooling below X-ray emitting temperatures than is expected from the X-ray luminosity \citep{Peterson01,Peterson03,Boehringer01,Peterson06,McNamara07}. The implication is that the intracluster medium (ICM) must be heated.

On the other hand, large quantities of cool gas and dust have been found in brightest cluster galaxies at the centre of cool-core clusters. Searches conducted at optical and infrared wavelengths reveal both warm and cool gas reservoirs (e.g. \citealt{Edge01}; \citealt{Salome03}; \citealt{Hatch05}; \citealt*{Jaffe05}; \citealt{Johnstone07}). The BCGs  in cool-core clusters generally have anomalous blue colours implying recent star formation, and emit in recombination and low-ionization lines \citep{McNamara97,Peres1998,Cavagnolo2008}. These observations imply a relationship between the properties of the cluster and the central galaxy, but this interaction has not yet been understood or quantified precisely.

RXCJ1504.1-0248 is one of the most massive cool core clusters, with a prominent central X-ray brightness peak, and a short central cooling time. A classical cooling flow model leads to a mass deposition rate of $1400-1900$\,\Msunpyr, and the brightest cluster galaxy emits strong low ionization emission lines  \citep{Boehringer1504}. These properties mark this cluster as a prime target to study the relationship between the ICM, the brightest cluster galaxy, and a possible central active galactic nucleus (AGN).

Here we report results of a multi-wavelength study of the core of RXCJ1504.1-0248. It lies at $z=0.2153$, which allows for a detailed spatial analysis of the BCG. We present results of integral field unit (IFU) spectroscopic observations aimed at understanding the morphology, kinematics and ionization state of the line-emitting nebula surrounding the BCG. We also present an analysis of X-ray spectra obtained with the XMM-Newton Reflection Grating Spectrometer (RGS) and European Photon Imaging Cameras (EPIC) and compare the ICM mass deposition rate to the star formation rate of the BCG. Sections \ref{sec:obs} and \ref{sec:dr} describe the observations and the data reduction process, while in section \ref{sec:results} we  present the properties of the BCG, the nebula, and the cluster core cooling rate. In section \ref{sec:disc} we combine the results from the multi-wavelength observations to quantify the relationship between the ionized gas, the BCG and the cluster core. We summarize our findings in section \ref{sec:concl}.

Unless otherwise stated, the following cosmological parameters have been used: $\Omega_m=0.3$, $\Omega_\Lambda=0.7$, $H_0=70\,{\rm km\, s^{-1}\, Mpc^{-1}}$. For a redshift $z=0.2173$ (the redshift of the BCG), the scale is $3.519\,\,{\rm kpc\,arcsec^{-1}}$.

\section{Observations}
\label{sec:obs}
\subsection{Optical images}
We observed RXCJ1504.1-0248 on 2006 March 1 with the Visible Multi-Object Spectrograph \citep[VIMOS;][]{LeFevre03} on the 8.2-m UT3 of the Very Large Telescope (VLT) at ESO Paranal in Chile, operated in imaging mode, using the R-band filter. The VIMOS field of view consists of four optical channels (also referred to as ``quadrants''), which in imaging mode cover a field of view of approximately $7'\times 8'$. The CCD size is $2048\times 2440$ px$^2$, therefore yielding a pixel size of 0.205 arcsec. Twelve scientific exposures were taken, each with an integration time of $\sim$ 90 s.

\subsection{Optical integral field spectroscopy}
On 2006 April 5-6, the central galaxy of the cluster was also observed with VIMOS in Integral Field Unit (IFU) mode with the LR-red and LR-blue grisms. Both grisms provide a ${\rm 54\times 54-arcsec^2}$ field of view, divided into 6400 optical fibres that offer a spatial sampling of 0.67 arcsec/fibre. The fibre to fibre distance on the sky is about 1 arcsec and the dead space between fibres is below $\sim$0.1 arcsec. Thus, the optical fibres, coupled to an array of 6400 microlenses, ensure a nearly continuous sky coverage. All quadrants of the spectrograph are divided into four pseudoslits, each organized into a ${\rm 20\times 20 }$ array of microlenses and holding five ${\rm 20\times 4}$ fibre modules.

The LR-red grism, used together with the OS-red filter, samples the wavelength range $5900-9150$ \AA, with a dispersion of approximately 7.3 ${\rm \AA \, pixel^{-1}}$ and a spectral resolution of 260. The LR-blue grism, used together with the OS-blue filter, covers the wavelength range $4000-6700$ \AA, with a dispersion and spectral resolution of approximately 5.3 ${\rm \AA \, pixel^{-1}}$ and 220, respectively. Eight 450-second exposures were taken in each setting, at averaged airmasses in the range $1.093-1.717$, and with a pointing dither of about $\left(\Delta\alpha, \Delta\delta\right) = \left(+3\,{\rm arcsec,\,} -3\,{\rm arcsec} \right)$. A summary of these observations is provided in Table \ref{tab:ifu}.

\begin{table}
  \centering
  \begin{minipage}{190mm}
  \caption{Summary of the VIMOS IFU observations.\label{tab:ifu}}
  \begin{tabular*}{0.44\textwidth}{@{}lllll@{}}
  \hfill
  &     & Wavelength & Exposure & Seeing \\
   Grism & Filter & coverage (${\rm \AA}$) & time (sec) & (arcsec) \\
 \hline
LR-red & OS-red & $5900-9150$ & $8\times 450$ & $0.5-1.3$ \\
LR-blue & OS-blue & $4000-6700$ & $8\times 450$ & $0.5-0.8$ \\
\hline
\end{tabular*}
 \end{minipage}
\end{table}

\subsection{X-ray observations}
RXCJ1504.1-0248 was furthermore observed with XMM-Newton on 2007 January 22, for 39 ks. 
We extracted a lightcurve for each of the EPIC detectors separately and excluded the time periods in the observation when the count rate deviated from the mean by more than 3$\sigma$ in order to remove flaring from soft protons \citep{Pratt02}. We did not find significant time intervals affected by flares, such that after this cleaning, the net effective exposure was $\sim$37.4 ks and $\sim$ 28.5 ks for EPIC/MOS and EPIC/pn, respectively.

\section{Data reduction}
\label{sec:dr}

\subsection{VIMOS Imaging}
\label{ssec:vimos_ima}

The imaging frames were reduced with the VIMOS Pipeline Recipes package, version 1.0, following all the steps described in section 5.5 of the VIMOS Pipeline User's Guide\footnote{ftp://ftp.eso.org/pub/cpl/vimos/vimos-pipeline-manual.pdf}, including flat fielding and bias subtraction. Flux calibration was applied using observations of standard stars in the L98 field\footnote{http://www3.cadc-ccda.hia-iha.nrc-cnrc.gc.ca/community/STETSON/standards/} taken on the same night as the science frames.

\subsection{VIMOS IFU}
\subsubsection{Reduction of the IFU data cubes}
The data reduction process was performed using the VIMOS Interactive Pipeline and Graphical Interface version 1.3 \citep{Scodeggio05}. Both the red and the blue datacubes underwent the usual VIMOS IFU reduction steps \citep{Zanichelli05,Covone06}, including adjustment of the ''first guess'' parameters of the instrumental model in order to correct for optical and spectral distortions, spectra extraction, cosmic-ray cleaning, wavelength calibration and flux calibration. The LR-red data were calibrated using the standard star EG-274, observed immediately after the science observations were taken. Unfortunately, no standard star was observed at the same time as the observations taken with the LR-Blue grism. Therefore, the flux calibration of the LR-blue spectra was done using observations of the standard star LTT-3864 taken on 2006 May 27.

Each fibre of the IFU has a slightly different transmission. To obtain accurate fluxes we corrected for different transmission between fibers. We measured the flux in the 7246 \AA\ sky\ line in the red data and 5577 \AA\ sky line in the blue data, and the spectrum from each fibre was normalized so that the flux in the sky line equals the median of the sky line flux from all the fibres. At this stage, spectra from fibres with low transmission were removed as they were too noisy. 

A sky spectrum was created for each fibre from the median of 20 nearby spectra in the red frame (or 30 fibres in the blue frame) which contained little or no line emission. These sky spectra were then subtracted from the calibrated spectra. Finally the spectra were trimmed to the wavelength range  3885--6225\AA\  and 5635--8348\AA, for the blue and red gratings respectively.

The absolute flux calibration of the IFU data after the VIPGI reduction is uncertain as our total flux measurement of [O{\sc ii}] is a factor of $\sim2$ lower than the slit spectrum of \citep{Boehringer1504}. Therefore we flux calibrate the IFU data using all stars and galaxies of known magnitude within the field-of-view  We find the SDSS magnitudes of these objects are consistent with those measured from our VIMOS image.  The IFU spectra of the stars and galaxies are convolved with g', r' and i' and VIMOS R-band filters, and the flux calibration is determined using SDSS magnitudes. This self-consistent calibration results in spectra which are 2.5 times brighter than spectra calibrated using VIPGI.

To increase the signal-to-noise ratio the spectra were binned using a Voronoi-2D binning algorithm\,\footnote{using {\sc idl} codes from http://www-astro.physics.ox.ac.uk/$\sim$mxc//idl/\#binning} \citep{Cappellari2003}. We use the weighted Voronoi tessellation as laid out by \citet{Diehl2006}  to have a signal-to-noise ratio of 15 in the H$\beta$ line flux measured in the blue frame.

\subsubsection{Spectral emission line fitting}
The emission lines were measured by fitting a linear continuum plus Gaussian model profile to each line using the {\sc idl} package {\sc MPFIT}~\footnote{http://cow.physics.wisc.edu/$\sim$craigm/idl/fitting.html}, which performs a Levenberg-Marquardt least-squares fit. 

The doublet [O{\sc ii}] $\lambda\lambda 3727,3729$ is never resolved, therefore it is fit with a single Gaussian. The H$\alpha$ line and the nearby [N{\sc ii}] $\lambda\lambda 6548,6583$ doublet are resolved and the complex is fitted with three Gaussians, with the integrated flux of the [N{\sc ii}] $\lambda 6548$ line fixed to be a third of the [N{\sc ii}] $\lambda 6584$ line. Similarly [O{\sc i}]$\lambda$6363 is fixed to be a third of the flux of  [O{\sc i}]$\lambda$6300, and  [O{\sc iii}]$\lambda$5049 is fixed at a third of  the flux of [O{\sc iii}]$\lambda$5007.

In the red wavelength domain, the spectra are strongly affected by fringing beyond 7000\AA. Consequently the signal-to-noise ratio at the wavelength of H$\alpha$ is not as high as that achieved at the wavelength of [O{\sc ii}].

\subsubsection{Extinction along the line of sight}
\label{ssec:ext}

We assume that there are two sources of dust which lead to extinction and reddening of the spectra. The first source is a screen of foreground dust at the redshift of the BCG and the other source is dust within our Galaxy. We first correct the spectra for Galactic extinction of E(B-V)=0.108 \citep{Schlegel1998}, then blue-shift the spectra to the rest-frame of the BCG, and correct for extinction caused by dust in the BCG. 

The extinction caused by dust within the BCG is determined from the observed  H$\alpha$/H$\beta$ ratio. The H$\alpha$/H$\beta$ flux is a good indicator of extinction along the line of sight, as shorter wavelengths are scattered stronger by dust particles. Therefore, in dusty regions, values of H$\alpha$/H$\beta$ are significantly above the theoretical ratio of 2.87 derived by \citet{Osterbrock06} for Case B recombination at $T_{\rm e}=10,000\,{\rm K}$. We assume the H$\beta$ flux is depleted due to 2\AA\ of H$\beta$ stellar absorption \citep[the mean found in a sample of star forming galaxies,][]{Buat2002} and we correct the H$\alpha$/H$\beta$ ratio for this before calculating the extinction. 


\begin{figure}
  \begin{center}
 \includegraphics[width=1\columnwidth]{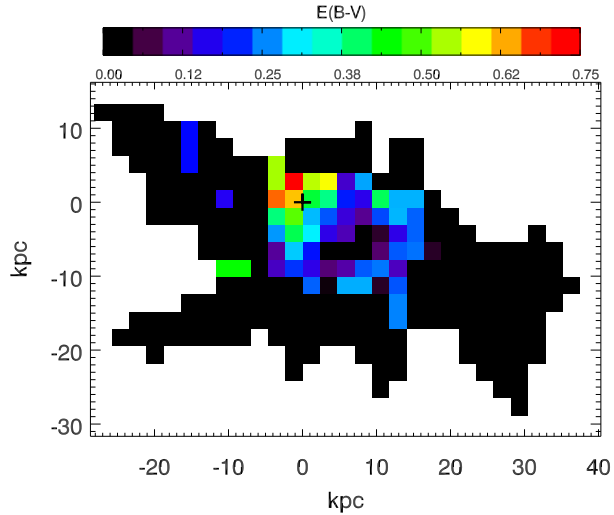}
\caption{Extinction map [$E$($B-V$)] of RXJ\,1504 derived from the ratio of H$\alpha$/H$\beta$ measured in each bin. The centre of the galaxy is marked by a cross. Black bins indicate areas where [O{\sc ii}] is detected, but H$\beta$ is either not detected or the H$\alpha$/H$\beta$ ratio implies $E$($B-V$)=0. \label{fig:ebv}}
\end{center}
\end{figure}

The $E(B-V)$ colour excess in each bin was computed using the equations from \citet*{MKT06},
 \begin{eqnarray}
  E({\rm H}\beta-{\rm H}\alpha) &\equiv & -2.5\,{\rm log} \left[\frac{\left({\rm H}\alpha/{\rm H}\beta\right)_{\rm theo}}{\left({\rm H}\alpha/{\rm H}\beta\right)_{\rm obs}}\right] \\
  E(B-V) &\equiv& \frac{E\left({\rm H}\beta-{\rm H}\alpha\right)}{k_{{\rm H}\beta}-k_{{\rm H}\alpha}} \,,
 \end{eqnarray}
 where $k_{{\rm H}\beta}-k_{{\rm H}\alpha}=1.278$ for R$_{\rm V}$=3.2. The spatial distribution of the extinction is plotted in Fig. \ref{fig:ebv}.  The central region of the galaxy is dustier than the outskirts, but the dust is patchy and there is no overall smooth pattern. 
The spectra in each bin was blue-shifted to the rest-frame and then corrected for extinction using the equation of \cite{Calzetti00} with the R$_{\rm V}=3.2$ extinction law.

\subsection{Reduction of the XMM-Newton data}
\label{ssec:xmm}

We reduced the {\it XMM-Newton} data using the 7.0.0 version of the Science Analysis System (SAS); the standard analysis methods using this software are described in e.g. \citet{Watson01}. 

We subtracted the instrumental background from the EPIC spectra using closed-filter observations which were normalized to match our observation using the count rates in the hard energy band ($10-12$ keV for MOS, $12-14$ keV for pn) outside of the field of view (OoFoV). Out-of-time events were subtracted from the EPIC/pn data using the standard SAS prescription for the extended full frame mode. The cosmic X-ray background (CXB) obtained from spectra extracted from an annulus between 9.5\arcmin\ and 12.5\arcmin\ (outside the cluster's $r_{500}$) is modeled with three components: two thermal components to account for the local hot bubble (LHB) emission ($kT_1 = 0.08$ keV) and for the Galactic halo (GH) emission ($kT_2 \sim 0.2$ keV), as described by \citet{kuntz2000}, and a power-law to account for the integrated emission of unresolved point sources and for possible contamination from the residual soft proton particle background. Since different detectors can be affected differently by soft protons, we leave the power-law indices and normalizations free between EPIC/MOS1, MOS2 and pn. The temperature of the LHB was frozen to 0.08~keV, while the temperature of the GH and the spectrum normalizations of the thermal components were free in the fit, but constrained to be the same for all three EPIC detectors. The spectra were fit in the 0.4--7. keV energy band.

The RGS spectra were extracted following the method described by \citet{tamura2001b}.
We modeled the background using the standard background model available in SAS \citep[{\texttt{rgsbkgmodel}},][]{riestra2004}.
The cluster spectra were extracted from a region which is 2\arcmin\ wide in the cross-dispersion direction of the instrument.
The line emission observed with the RGS from extended sources is broadened by the
spatial extent of the source along the dispersion direction.
In order to account for the line broadening in the spectral modeling, we convolve
the line spread function (lsf) model with the surface brightness profile of the
source along the dispersion direction derived in the 0.8--1.4~keV band.
We fit the 1st order spectra in the wavelength band of 8--25~\AA\ and the second order spectra in a narrower band of 8--16~\AA.
The 1st and 2nd order spectra obtained by the two RGS detectors were fit simultaneously with the relative instrument normalizations left as free parameters.

Freezing the Galactic absorbing column density ($n_{\rm H}$) to $6\times10^{20}\:\rm{cm}^{-2}$, as determined from \citet{dickey1990} from 21 cm observations, gives very poor fits to the cluster EPIC spectra. Therefore, we froze the $n_{\rm H}$ to $8.7\times10^{20}\:\rm{cm}^{-2}$, the best-fit value determined by fitting EPIC spectra of the central $5^\prime$ in combination with ROSAT All-Sky Survey (RASS) spectra to better constrain the emission in the soft X-ray band. This value agrees within the 90\% confidence interval with the $n_{\rm H}$ value computed from the 100 $\mu$m IR data, using the \citet{Boulanger96} IR-$n_{\rm H}$ correlation function (more details will be provided by Moi\c{s} et al., in prep).

\section{Results}
\label{sec:results}

\subsection{The mass deposition rate from X-ray spectra}
\label{ssec:mdr}

To determine the amount of ICM gas cooling out of the X-ray emitting temperature range, we extracted the EPIC spectrum from a circle of radius 140 kpc (0.67\arcmin) centred on the cluster centre. This is the cooling radius obtained by \cite{Boehringer1504} for a Hubble constant $H_0$ of 70 km s$^{-1}$ Mpc$^{-1}$. 

The spectrum is fitted with a Galactically absorbed single temperature collisionally ionized optically thin {\it mekal} plasma model plus a classical cooling flow model (with the lower cutoff temperature fixed to 0.1 keV). To ensure a correct propagation of errors due to uncertainties in background determination, the EPIC source spectra are fit in parallel with the CXB spectra obtained from an outer annulus. The normalizations of the cluster spectral components were set to zero for the CXB data sets, while the normalizations of the background models were fixed for the source analysis to a ratio corresponding to the relative sizes of the extraction regions for the source and CXB only. 

The metal abundances were coupled between the single temperature and cooling flow models both for fitting the EPIC and the RGS data. The O/Fe, Ne/Fe and Mg/Fe ratios were fixed in the EPIC fit based on the best-fit values obtained from the RGS spectra, since the energy resolution and effective area calibration of EPIC around the energy of the O, Mg and Ne lines are too poor to allow a reliable determination of their abundances with EPIC alone, especially for such a hot cluster. The Si and S abundances of the hot plasma can only be measured with EPIC because the emission lines of these elements lie outside the RGS energy band. The fit results are summarized in Table \ref{tab:rgs}. Errors are quoted at the $1\sigma$ level. Abundance ratios are given in proto-solar units \citep{Lodders}.

\begin{table}
\caption{XMM-Newton spectral fit results for the central region of RXCJ\,1504.1-0248. The abundance values are given with respect to solar abundances \citep{Lodders}. Errors are quoted at the 1$\sigma$ level.}
\begin{center}
\begin{tabular}{l|cc}
\hline
& RGS & EPIC\\
\hline
k$T$ (keV)      & $8.8^{+1.3}_{-1.0}$    &   $5.6^{+0.06}_{-0.05}$    \\
O/Fe               & $0.47\pm0.25$ &  0.47 (fixed)  \\
Ne/Fe              & $0.71\pm0.35$  & 0.71 (fixed)   \\
Mg/Fe              & $0.57\pm0.32$  &  0.57 (fixed)  \\
Si/Fe           & --  & $0.79\pm0.12$    \\
S/Fe          & --   & $0.53\pm0.15$     \\
Fe              & $1.77^{+0.94}_{-0.56}$ & $0.47^{+0.01}_{-0.01}  $    \\
$\dot{M} \: {\rm (M_{\sun}/yr)}$  & $68^{+57}_{-56}$ & $78^{+26}_{-19}$ \\
\hline
$\chi^2$ / d.o.f. & 467/481  &  3003/2494     \\
\hline
\end{tabular}
\label{tab:rgs}
\end{center}
\end{table} 

\begin{figure}
\includegraphics[height=\columnwidth,angle=270]{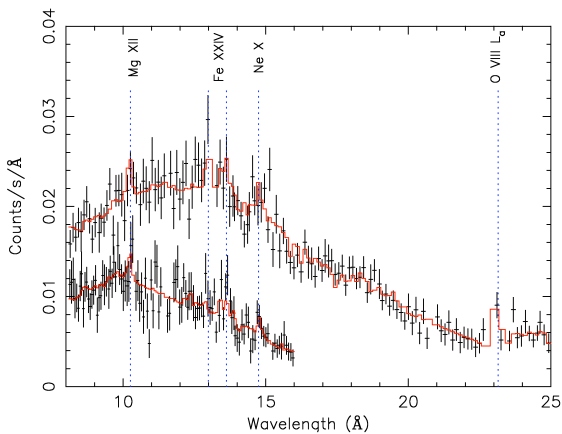}
\caption{Combined RGS 1st and 2nd order spectrum of RXCJ1504.1-0248 and best-fit model.}
\label{fig:rgs}
\end{figure}

The high luminosity and the highly peaked surface brightness distribution of the cooling core in RXCJ1504.1-0248 allow us to obtain relatively sensitive XMM-Newton RGS spectra of this cluster despite its large distance. We show in Fig. \ref{fig:rgs} the spectrum obtained by combining data from the two RGS detectors. This is one of the most distant galaxy cluster spectra obtained with RGS. The cause for the discrepancy between the temperatures measured with EPIC and RGS is the different spectral extraction region (the RGS extraction region is effectively 10\arcmin-long in the dispersion direction and 2\arcmin-wide in the cross-dispersion direction, and the temperature profile of the cluster rises steeply outside the cooling radius of 0.67\arcmin). Because of the higher temperature measured by RGS, the measured absolute abundances are also high, but the abundance ratios are typically robust. 

The best-fit values of the mass deposition rates from the EPIC and RGS data are 78 and 68 ${\rm M_{\sun}\:yr^{-1}}$ respectively. The 3$\sigma$ upper limits are 156 and 239 ${\rm M_{\sun}\:yr^{-1}}$, respectively.

The so-called ``cooling flow problem'' remains in that the mass deposition rates obtained from X-ray spectra are significantly less than those inferred from the X-ray luminosity of between 1400 and 1900 ${\rm M_{\sun}\:yr^{-1}}$ for  $H_0$ = 70 km s$^{-1}$ Mpc$^{-1}$. This implies that significant heat input is needed to reduce the cooling from 1400--1900 down to at most 150--240 solar masses per year ($\sim$10\%). The bolometric luminosity calculated by \cite{Boehringer1504} is $4.3\times10^{45}$ ergs s$^{-1}$, out of which more than 70\% ($L_{Xc}=3\times10^{45}$ ergs s$^{-1}$) is emitted from inside the cooling radius. Assuming 10\% of the gas is cooling and forming stars, more than 90\% of $L_{Xc}$ must thus be supplied by a heat source, amounting to $2.7\times10^{45}$ ergs s$^{-1}$.

\subsection{Galaxies within the cluster core}
\label{sec:results:galaxies}
\begin{figure*}
 \begin{center}
\centering
 \includegraphics[width=1.0\columnwidth]{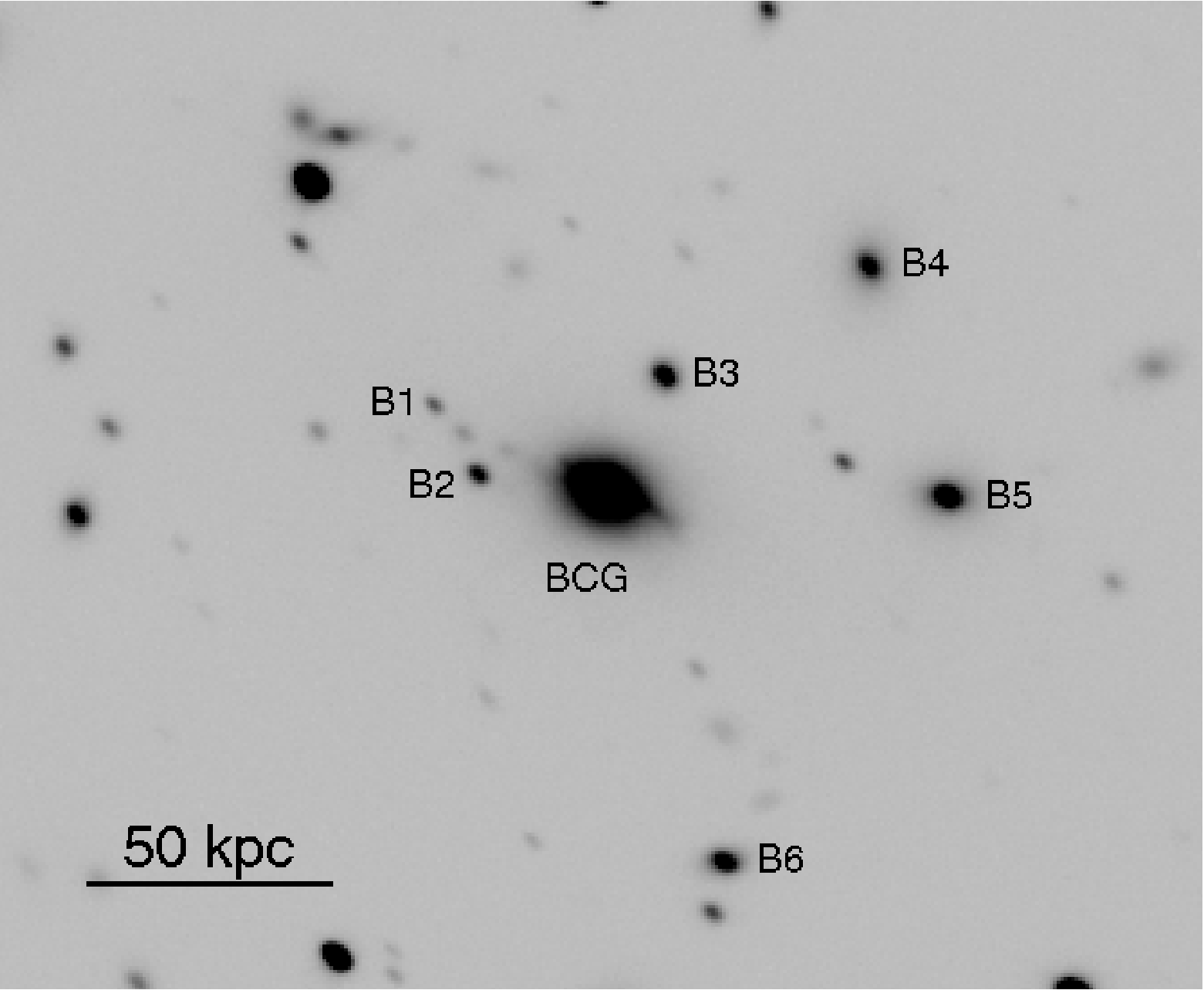}
 \includegraphics[width=1.0\columnwidth]{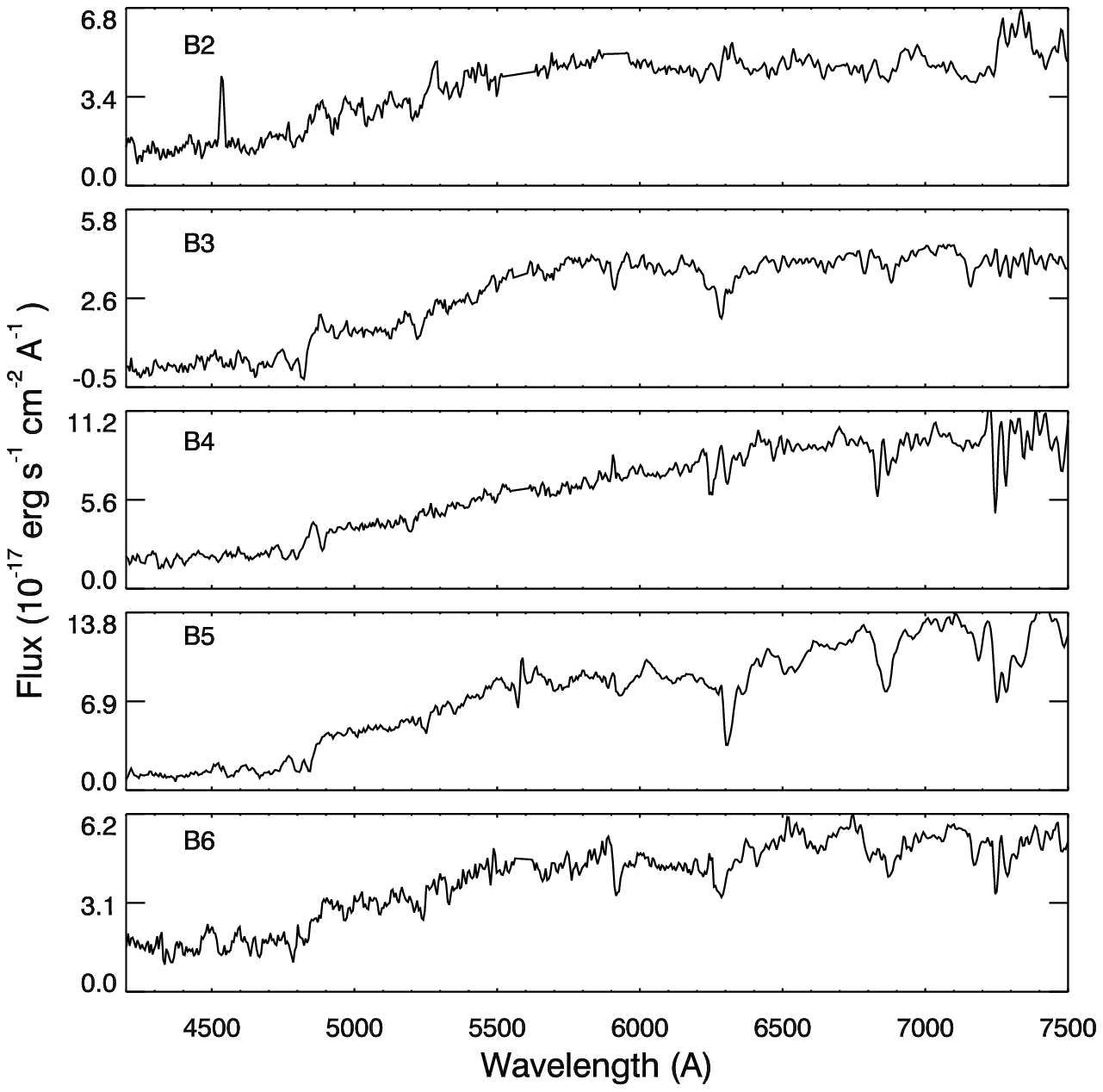}
 \caption{ {\bf Left:} $R-$band VIMOS image of the entire field-of-view of the VIMOS IFU centered on the brightest cluster
galaxy (labelled BCG). A number of other nearby bright galaxies have been labelled B1--B6. 
 {\bf Right:} Spectra of galaxies B2--B6. Most absorption features above 6000\AA\ are due to atmospheric absorption. Stellar absorption Ca K and H lines are detected in all galaxies and [O{\sc ii}] is seen in the spectrum of B2.
 \label{fig:VIMOS_continuum}}
 \end{center}
\end{figure*}

\begin{figure}
 \begin{center}

\includegraphics[width=1.0\columnwidth]{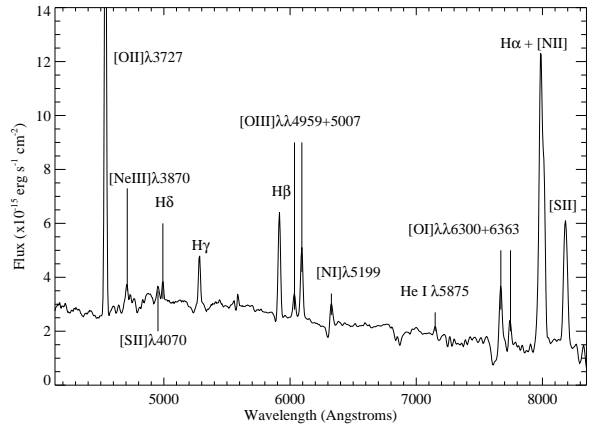}

 \end{center}
 \caption{Extinction-corrected composite spectrum of the ionized nebula surrounding the BCG. The full height of the [O{\sc ii}] emission line is not shown. Emission line ratios are given in Table\,\ref{tab:ratios}.\label{fig:spectrum}}
\end{figure}

The cluster core contains the BCG and a number of fainter galaxies marked B1--B6 in the VIMOS R-band image (Fig.\,\ref{fig:VIMOS_continuum}). Spectra of these galaxies are shown in Fig.\,\ref{fig:VIMOS_continuum} and the BCG spectrum in Fig. \ref{fig:spectrum}. The light from B1 fell on some bad IFU fibres so we were not able to extract a spectrum. All spectra were corrected for Galactic extinction of $E$($B-V$)$=0.108$ \citep{Schlegel1998} and the BCG was corrected for additional intrinsic extinction of $E$($B-V$)$=0.211$ derived from the measured H$\alpha$/H$\beta$ ratio of 3.68 (assuming 2\AA\ of H$\beta$ absorption). The BCG emits strongly in recombination and low ionization lines, and the observed and extinction-corrected emission-line ratios are given in Table \ref{tab:ratios}. 

Redshifts of galaxies B2--B6 were measured using the strong stellar absorption lines Ca {\sc ii} H and K near 4000\AA. These lines are not visible in the BCG spectrum so the [O{\sc{ii}}] emission line is used instead. Redshifts are listed in Table\,\ref{tab:galaxies}, together with the strength of the 4000\AA\ break (D4000) and the rest-frame $B-R$ colour. 

Galaxies B2, B3, B5 and B6 lie within a few hundred \kmps\ of the redshift of the BCG therefore these galaxies are situated within the cluster. B4 is blue-shifted by almost 2000\,\kmps\ compared to the BCG and is either a galaxy in the infall region of the cluster or a foreground galaxy.

D4000 is defined as the ratio of the average continuum between the rest-frame 4050--4250\AA\ and 3750--3950\AA\ \citep{Bruzual1983}. However, this wavelength range includes strong emission lines visible in the spectrum of the BCG. To remove emission-line contamination from the D4000 measurement, we alter our definition for the BCG to the ratio between rest-frame 4115-4250\AA\ and 3750-3840\AA . 

The strength of this break, D4000, is typically greater than $1.6$ in elliptical galaxies \citep{Kauffmann2003}, even if the elliptical galaxies reside at redshifts of up to $z\sim1.1$ \citep{Pasquali2006b}.  The cluster galaxies B2--B6 have D4000 values typical of ellipticals, but the BCG has a much lower D4000, similar to late-type galaxies. 

B2 is the only galaxy, apart from the BCG, to exhibit [O{\sc ii}] line emission. Although it lies close to the emission-line nebula that surrounds the BCG, the kinematics of the emission line gas (see section \ref{sssec:kin}) imply the line is emitted from gas within B2 and not from the BCG nebula.

\begin{table*}
  \begin{tabular}{l||cccccc}
  \hline
Galaxy&  BCG & B2 & B3 &B4&B5&B6\\ \hline
Redshift&0.2173&0.2176&0.2151&0.210&0.221&0.216\\
Velocity shift relative to BCG (\kmps)&--& 65& $-550$& $-1885$&  800& $-385$\\
Projected distance (kpc) &--& 24.6 &  26.3&  70.7& 68.8&85.5\\
D4000&1.28 (1.11)$^{2}$&1.77& 3.84& 1.79& 2.85 & 1.80\\   
$B-R$ colour$^{1}$ (mag)& $-0.12$ ($-0.40$)$^{2}$& 0.0&  0.13& 0.43&   0.43&  0.23\\ \hline
\end{tabular}
  \caption{Properties of the brightest cluster galaxy and galaxies nearby in projection. Redshifts are determined from the Ca {\sc ii} H and K absorption lines. Projected distances are given in kpc, as observed from the dust-uncorrected continuum. $B-R$ colours are measured from regions free from any emission line contamination. $^{1}$ The $B-R$ colour is the magnitude difference between  5600-5800\AA\ and  6900-7100\AA. $^{2}$ Values in the brackets are derived from the extinction-corrected spectrum.  \label{tab:galaxies}}
\end{table*}

\begin{table}
  \begin{tabular}{lcc}
  \hline
Line& Ratio to H$\beta$ & Extinction-corrected ratio\\
\hline
  
$[$O{\sc ii}]  &  3.00& 4.30\\
$[$Ne {\sc iii}] &0.10&0.14\\
H$\gamma$   &  0.41&0.48\\
H$\beta$   &1.00&1.00\\
$[$O {\sc iii}]$\lambda5007$ &0.7&0.68\\
$[$N {\sc i}]$\lambda5199$  &0.25&0.23\\
He {\sc i}$\lambda5785$    &0.11&0.09\\
$[$O{\sc i}]$\lambda6300$    &1.21&0.88\\
H$\alpha$    &4.37&3.05\\
$[$N {\sc ii}]$\lambda\lambda6548+6484$   &3.54&2.46\\
$[$S {\sc ii}]$\lambda\lambda6717+6731$  &2.88&1.95\\
\hline
\end{tabular}
  \caption{ Emission line ratios for the observed and the extinction-corrected spectrum of the BCG. The ratio of H$\alpha$/H$\beta$ is greater than 2.87 due to our assumption that the H$\beta$ emission line is depleted due to 2\AA\ of stellar absorption. \label{tab:ratios}}
\end{table}

\begin{figure*}
 \begin{center}
\centering
  \includegraphics[width=0.605\columnwidth]{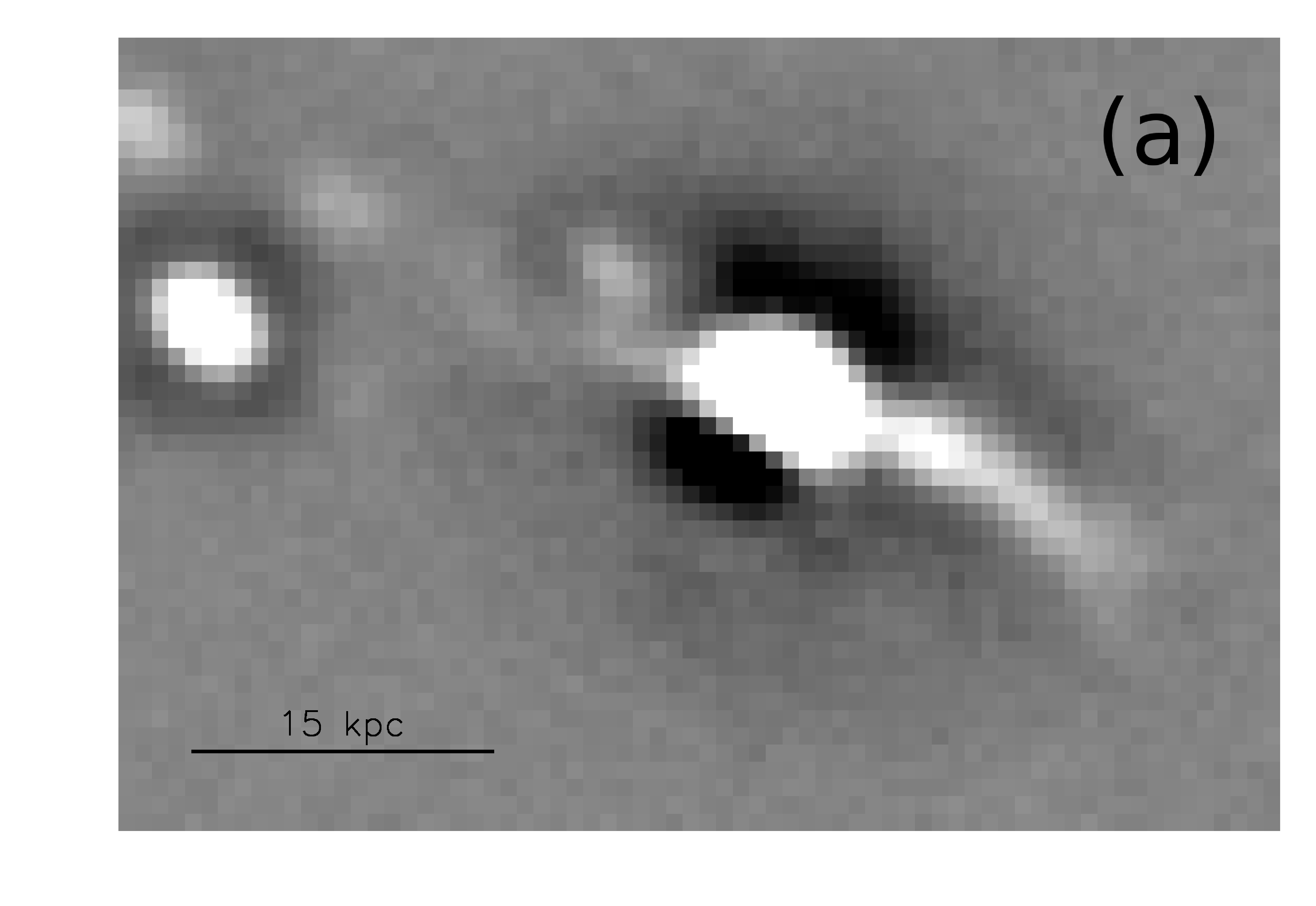}
  \includegraphics[width=0.66\columnwidth]{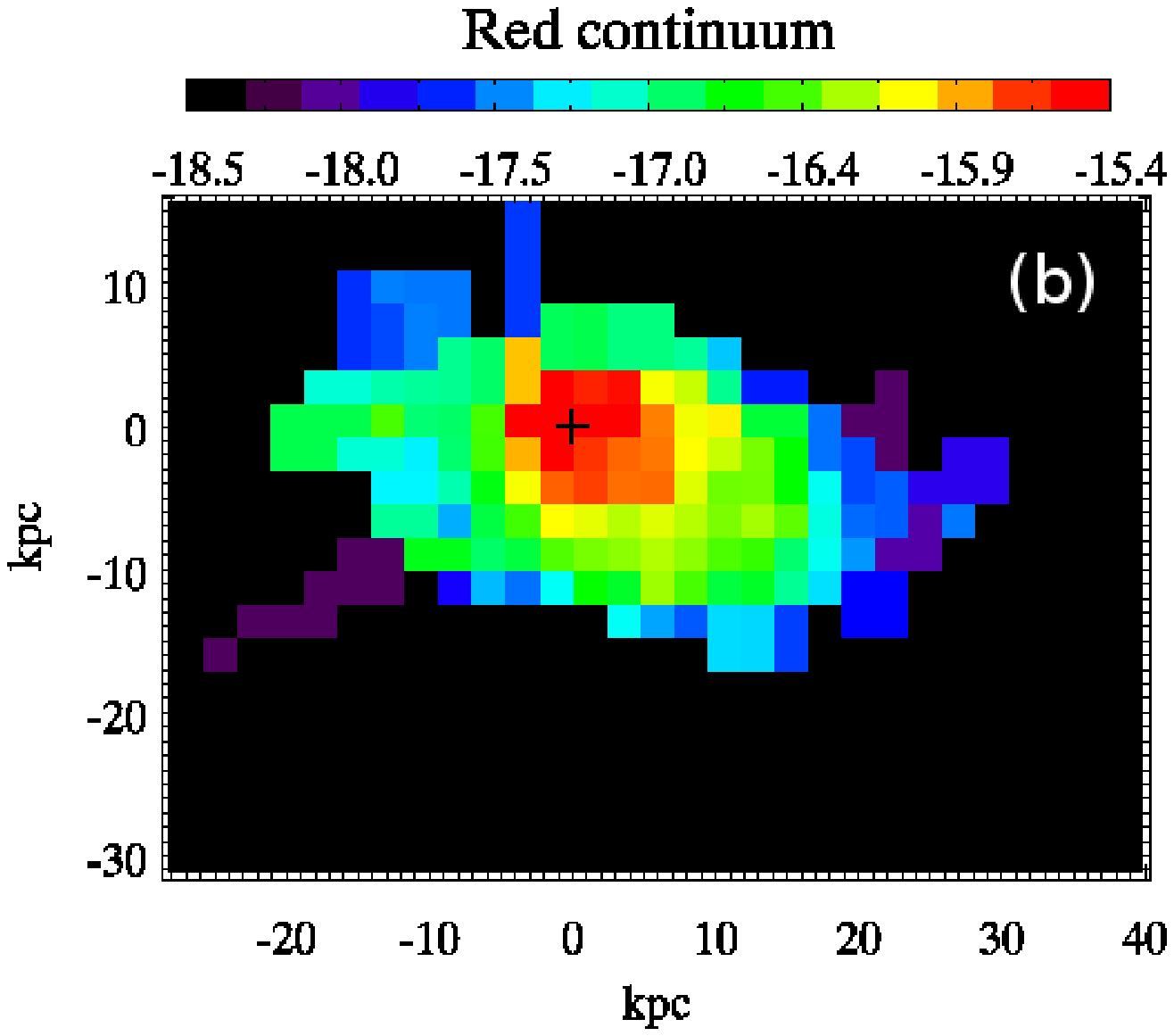}
  \includegraphics[width=0.66\columnwidth]{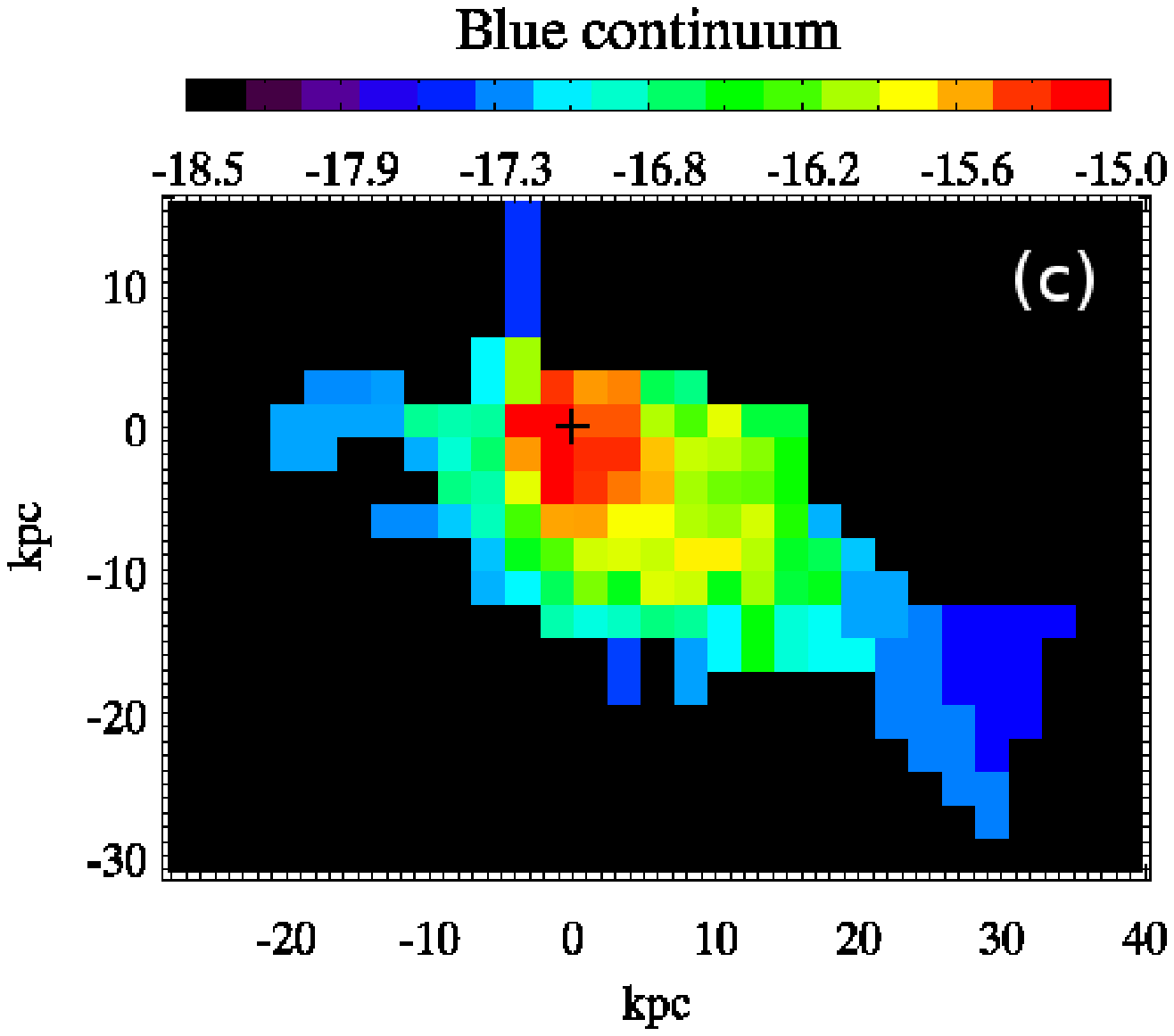}
  \includegraphics[width=0.66\columnwidth]{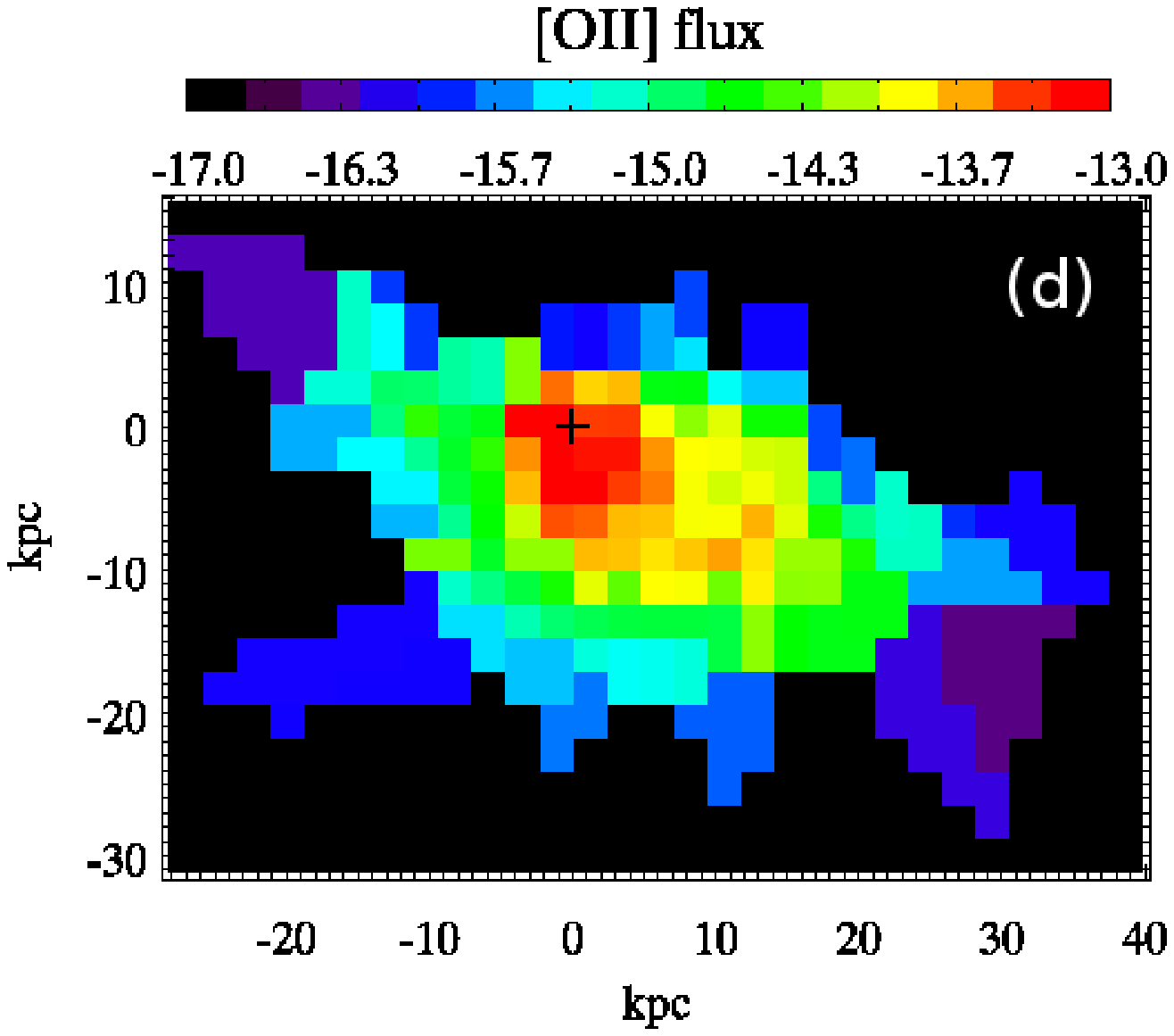}
  \includegraphics[width=0.66\columnwidth]{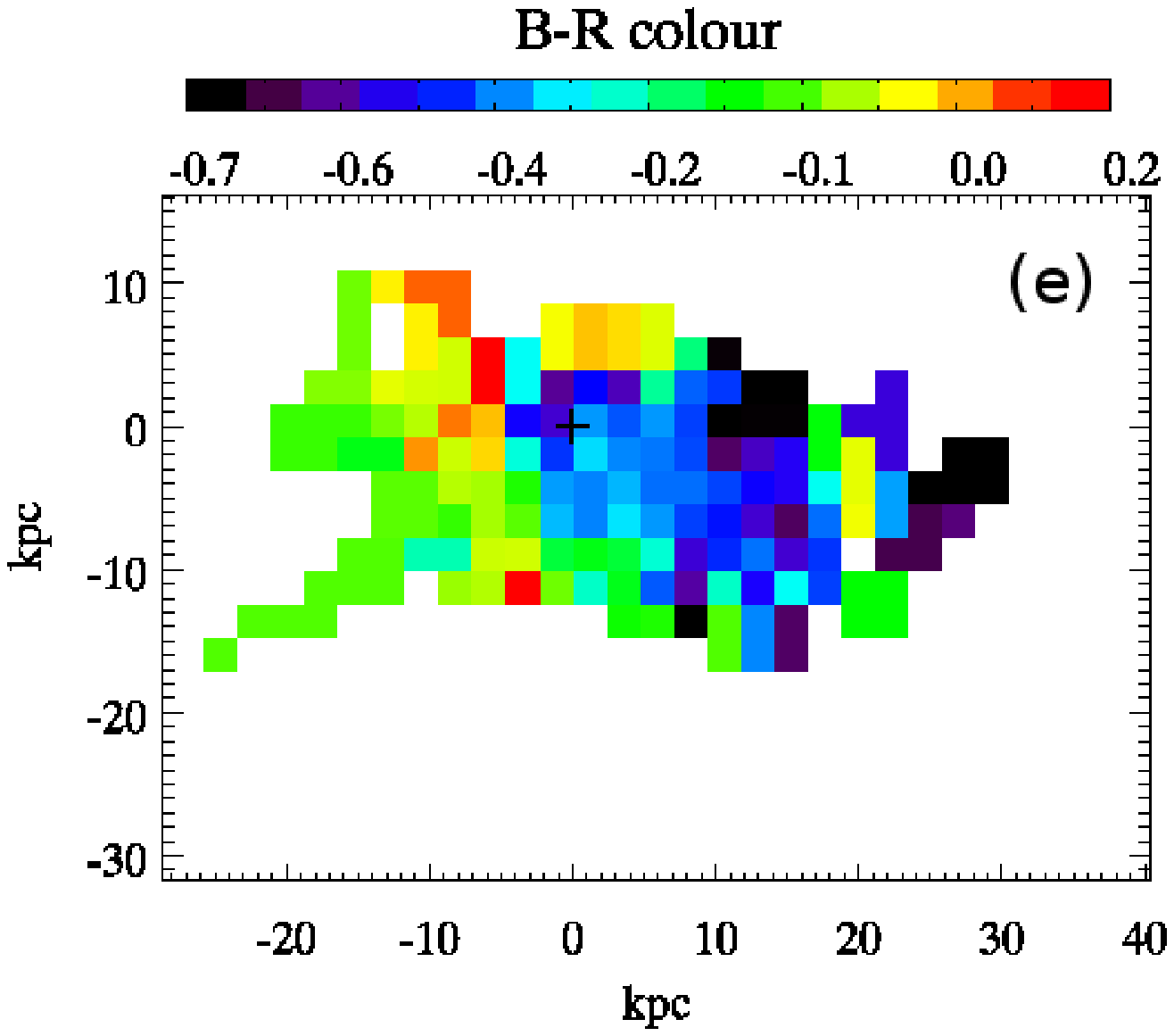}
  \includegraphics[width=0.66\columnwidth]{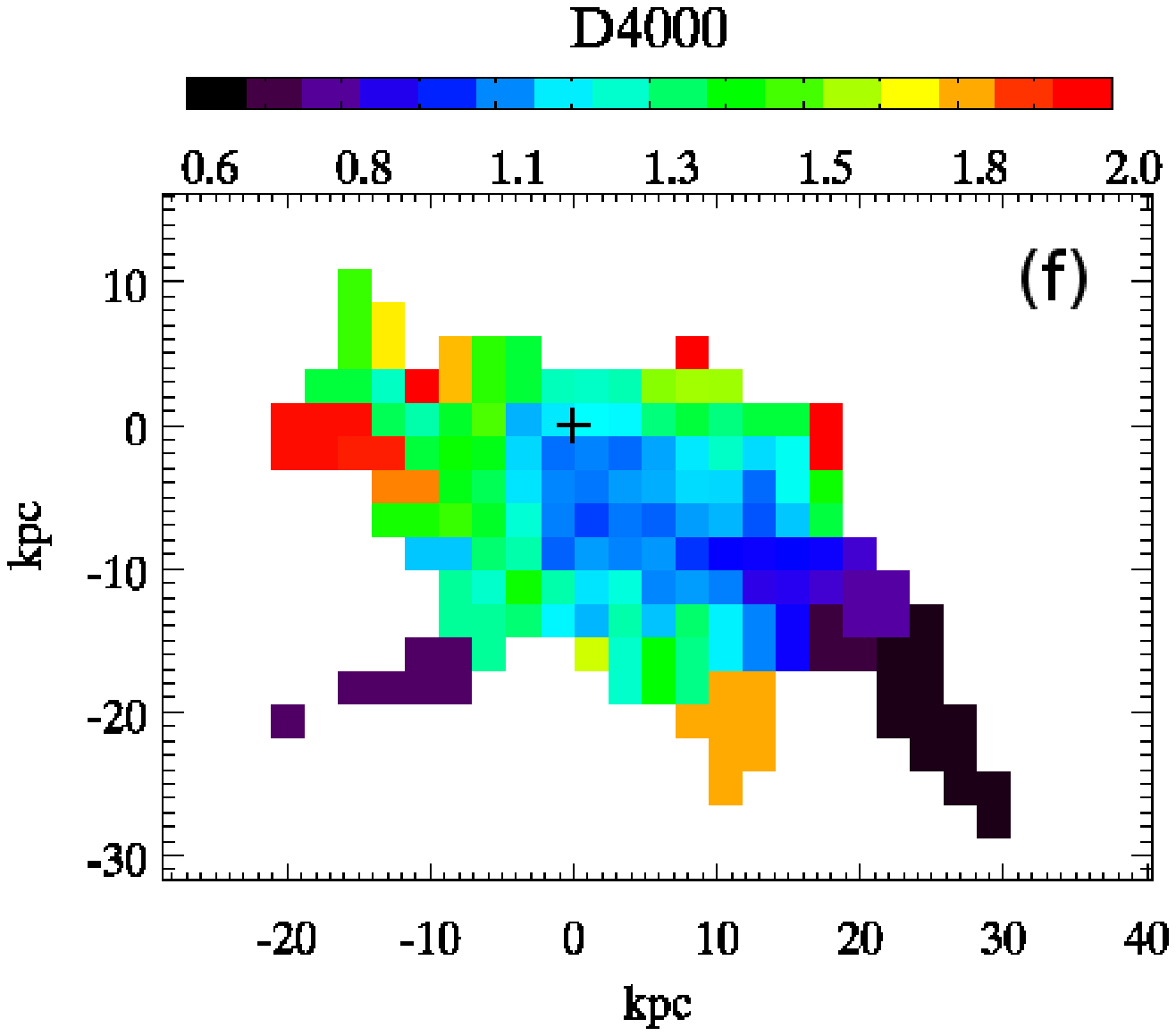}
  
 \caption{Continuum images of the BCG from left to right: (a) unsharp-mask of the $R-$band VIMOS image showing the NE and SW filament; (b) continuum from the rest-frame $\sim$5400\AA\ in log units of \ergpscmpspa arcsec$^{-2}$; (c) continuum from the rest-frame $\sim$3200\AA\ in log units of \ergpscmpspa arcsec$^{-2}$; (d) map of the [O{\sc ii}] emission with colourbar in log scale in units of  \ergpscmps arcsec$^{-2}$; (e) rest-frame $B-R$ continuum colour and (f) D4000. The noisy regions (where the flux per binned region is less than the rms noise) are coloured black in (b)-(d) and white in regions (e) and (f). The galaxy B2 is located at (-20,0) in the IFU images. 
  \label{fig:ifu_continuum}}
 \end{center}
\end{figure*}
\begin{figure*}
 \begin{center}
  \centering
  
   \includegraphics[width=0.66\columnwidth]{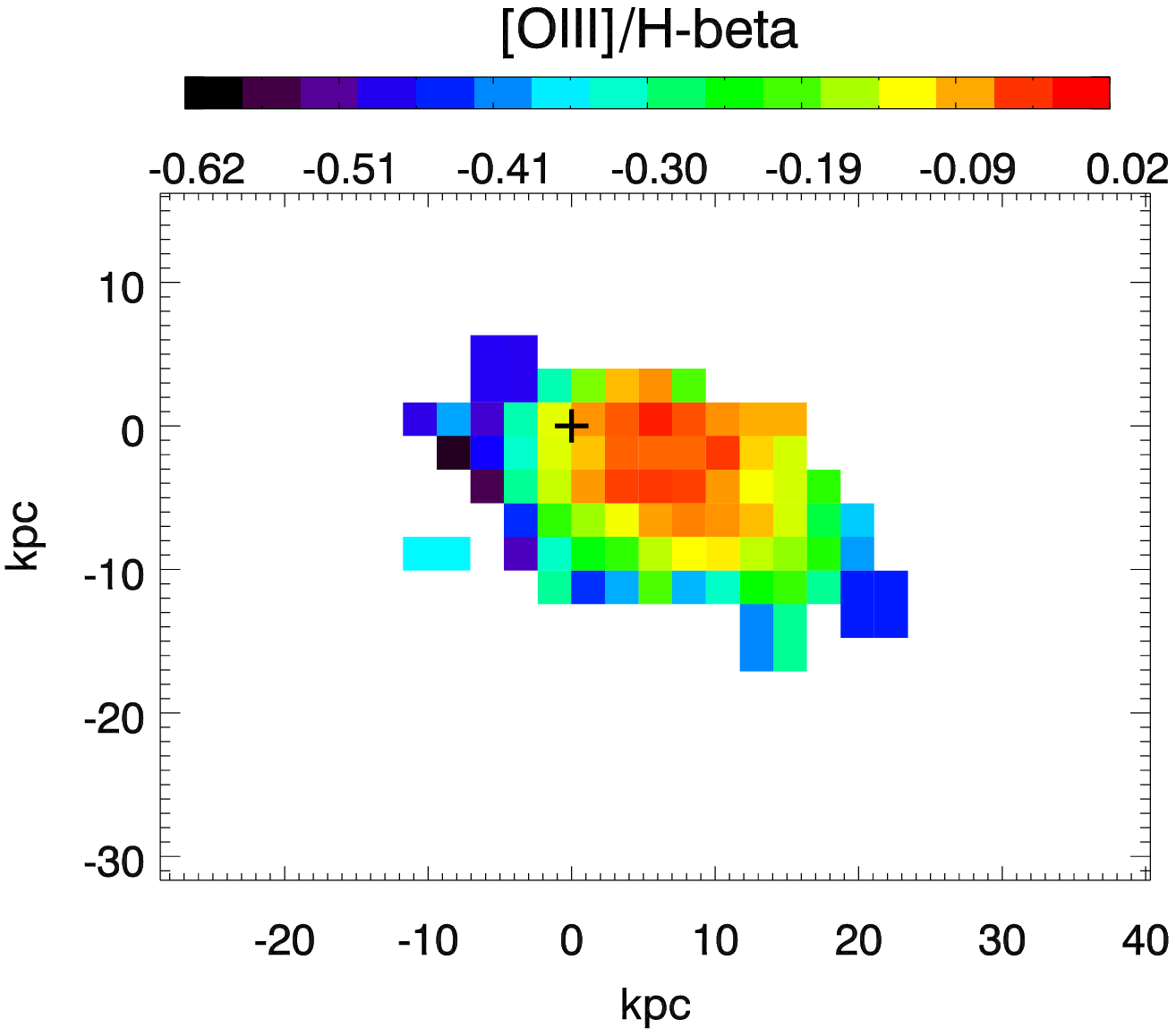}
   \includegraphics[width=0.66\columnwidth]{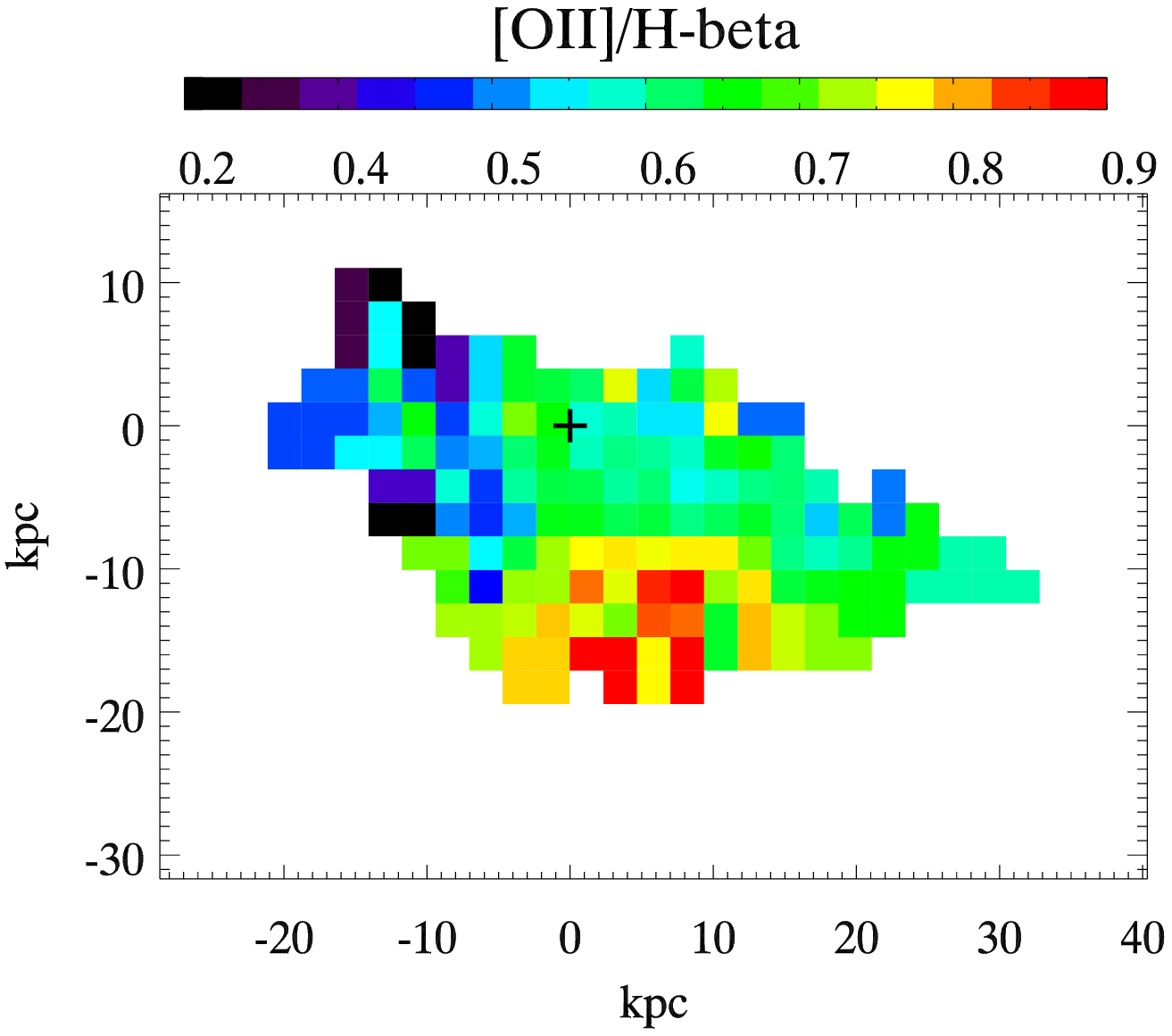}
   \includegraphics[width=0.66\columnwidth]{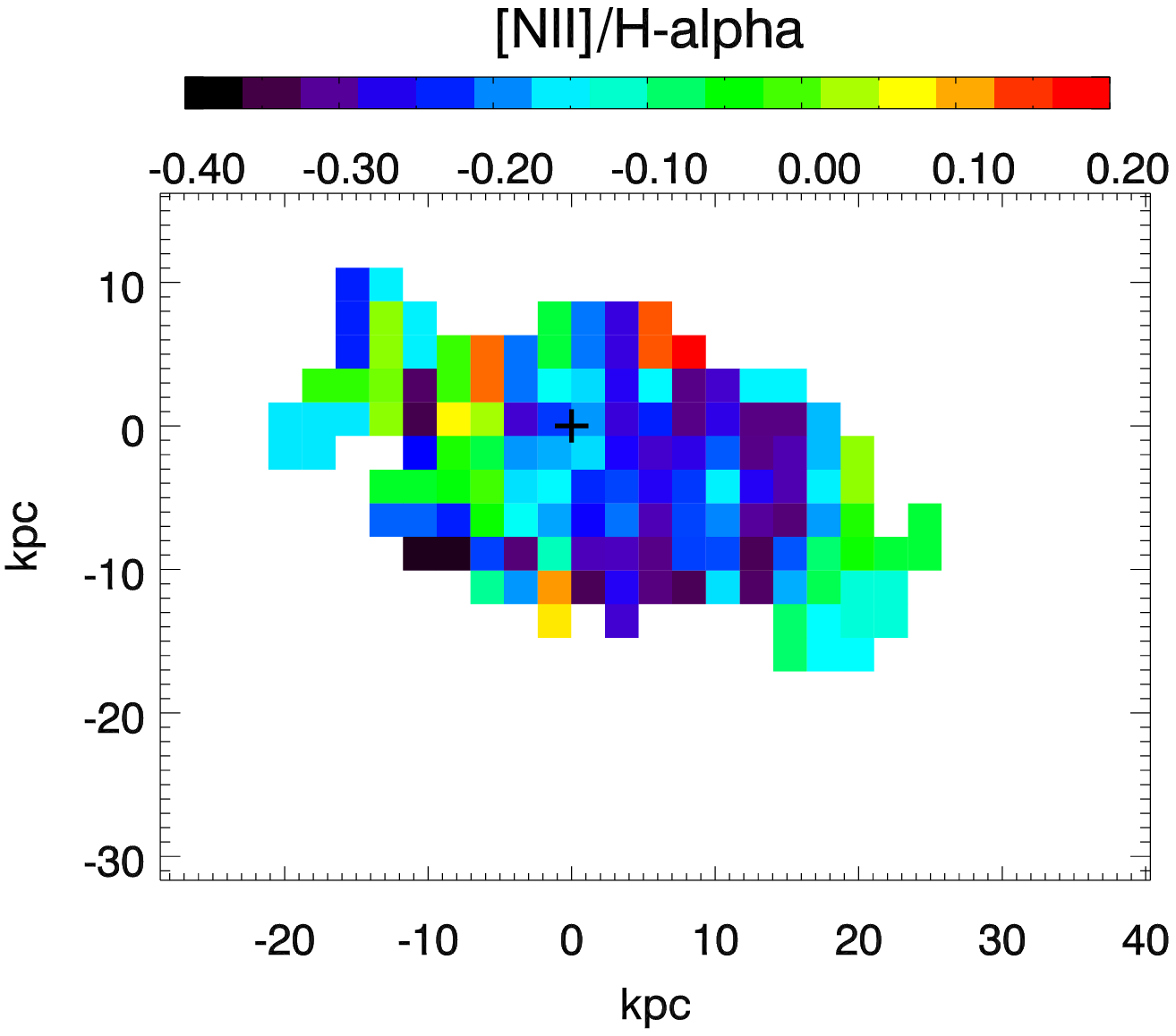}
   \caption{Emission line ratios of the nebula of the BCG from left to right: [O{\sc iii}]/H$\beta$, [O{\sc ii}]/H$\beta$, [N{\sc ii}]/H$\alpha$.   Binned regions are only displayed where [O{\sc iii}], H$\beta$, [N{\sc ii}]  and H$\alpha$ fluxes are greater than 5$\times10^{-17}$ \ergpscmps arcsec$^{-2}$. All scales are in log. \label{fig:ifu_lines}}
 \end{center}
\end{figure*}

\subsection{The brightest cluster galaxy}
\subsubsection{Mass}
The BCG of RXCJ1504.1-0248 is a massive galaxy: its $K_{\rm mag}$ is 13.130 (2MASS). The mass-to-light ratios of \citet{Cappellari2006} imply a stellar mass of $\sim6\times10^{12}$\Msun. This is twice the mass of the nearby cD galaxy M87. Although no K-correction has been applied, the galaxy is red at these wavelengths with $H-K=0.546$, so the K-corrected $K_{\rm mag}$ may be even larger. The dominant source of error in the derived mass is likely to be the large scatter in the mass-to-light ratio.

\begin{figure}
\includegraphics[width=1\columnwidth]{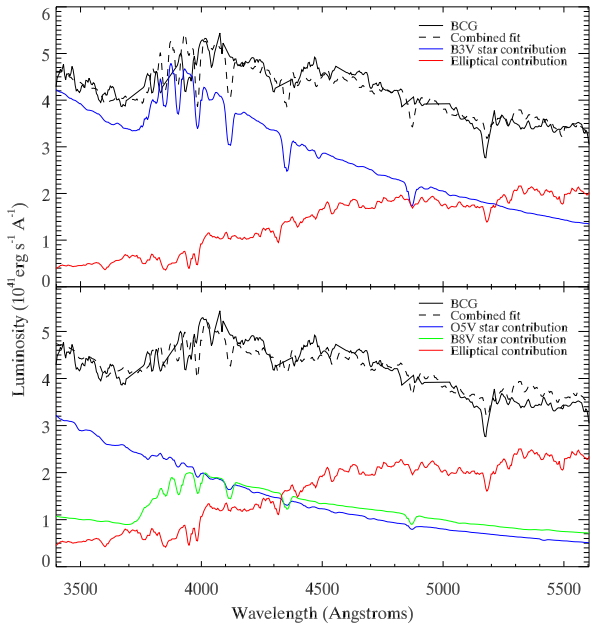}
\caption{Optical spectrum of RXCJ1504.1-0248 with emission-lines masked out (black solid line) and the best-fit models overplotted in the dashed line. The top panel shows a 2-component fit to the SED consisting of an elliptical galaxy template plus a B3V stellar template. The coloured lines show the relative luminosity from each component. The bottom panel shows a 3-component fit to the SED consisiting of an elliptical galaxy template, an O5V stellar template and a B8V stellar template. Both fits give similar reduced-$\chi^2$ and are thus equally likely.}
\label{fig:fit_continuum}
\end{figure} 

\subsubsection{Stellar populations}
\label{sec:stellarpop}
The BCG has a different spectral shape compared to the other cluster galaxies (B2--B6). Whereas the cluster galaxies have red $B-R$ colours and high D4000, typical of elliptical galaxies, the BCG has blue $B-R$ colours and shallow D4000.  These spectral features, in addition to the strong emission lines, means the BCG hosts a strong source of ultraviolet (UV) emission such as an AGN or a young stellar population.

We determine the stellar composition of the BCG by modeling the optical spectral energy distribution (SED).  Because the BCG is so massive we assume the underlying galaxy consists of an old stellar population which has the SED of an elliptical galaxy. The elliptical template from the Kinney-Calzetti Atlas \citep{Kinney1996} is used to model this component. 

The blue continuum was modeled by a power-law continuum (in which the slope and normalisation were free parameters) or early-type stellar templates  (O5V, O9V, B0V, B3V, B5-7V, B8V and A0V) from the Pickles Atlas \citep{Pickles1998}. The normalisation of the elliptical and stellar templates were free parameters. 

The emission lines were masked for the least-squared fitting procedure. Since the red part of the spectrum is contaminated by fringing, we only modeled the continuum between 3400--5600\AA\  and assign constant errors to the entire wavelength range of the optical SED. The errors were adjusted to obtain a reduced $\chi^2$ of approximately $\chi^2_{\rm red}\sim$1 for the best fit model.

The elliptical plus power-law model resulted in a poor fit to the BCG SED ($\chi^2_{\rm red}$=13) as the break between 3600-4000\AA\  was not adequately fit. Thus the blue continuum is not due to an AGN component.

The best fit to the data,  with a $\chi^2_{\rm red}$=1.8, resulted from the combination of the elliptical template and a B3V stellar template. This fit is shown in the top panel of Fig.\,\ref{fig:fit_continuum} together with the relative contributions of each component to the galaxy's luminosity. 

An elliptical plus O-star template is rejected with a  $\chi^2_{\rm red}$=6.5. The O-star template is similar to the AGN power-law SED, and the poor fit results from the same inadequate fit to the 3600-4000\AA\ break.

An O5V star emits 5 orders of magnitude more ionizing flux than a B3V star, thus to estimate the maximum ionizing flux allowed by the optical SED, we fit the optical SED with a 3 component model: an elliptical template, plus an O5V stellar template, and an additional stellar template from the O5V to A0V range.  The best fit to the data, with a  $\chi^2_{\rm red}$=1.5, results from the elliptical + O5V + B8V model. The bottom panel of Fig.\,\ref{fig:fit_continuum} shows the model overplotted on the optical SED of the BCG, and the relative contributions of each component. 

The difference in $\chi^2_{\rm red}$ between these two models results from the poor fit to the Balmer absorption lines. The observed SED contains strong emission lines so we are unable to fit any absorption lines to the data. The B3V stellar template has stronger absorption lines than the 3-component model and thus results in a larger $\chi^2_{\rm red}$. We therefore cannot distinguish which model is more plausible from the $\chi^2_{\rm red}$.

\begin{table*}
  \begin{tabular}{l cccccc}
  \hline
Model& \% E & \% O5V& \% B3V & \% B8V&no. of ionizing stars&H$\alpha$ flux \\
   &                    &             &              &            &$\times10^6$ & $10^{43}$erg\,s$^{-1}$\\ \hline
E + B3V&  37&--& 63& --&596&0.006\\
E + O5V + B8V&43& 27 &-- & 30 & 4.5 &3.4 \\ \hline
\end{tabular}
  \caption{Stellar populations in the BCG. The optical SED of the BCG is best fit by an old stellar population (modeled by an elliptical galaxy template, E), plus a large component of young stars having a B3V SED (E + B3V). In a 3-component fit which is forced to include an O5V stellar population, the SED is best fit by a composition of an elliptical galaxy, plus O5V and B8V stars (E + O5V + B8V). The percentage of light emitted at 4500\AA\ by each component is given in  columns \% E,\% O5V, \% B3V and \% B8V. The number of ionizing stars is the number of O5V or B3V stars, and column 7 lists the H$\alpha$ flux resulting from photoionization by these stars. The  E + O5V + B8V model  gives an upper limit on the H$\alpha$ flux produced by stellar photoionization.} \label{tab:stellarpop}
\end{table*}

In Table  \ref{tab:stellarpop} we list the relative flux contributions from the 2  and 3 component fits, and the number of ionizing stars (stellar types 05V and B3V) that lie in the BCG.  The monochromatic 4500\AA\ luminosities of an O5V and B3V star are $1.1 \times10^{34}$ and $4.8 \times10^{32}$\,erg\,s$^{-1}$, respectively \citep{Kurucz1993}. B8V and later stellar types that compose the elliptical galaxy template emit negligible amounts of ionizing photons. 

For both models approximately 60\% of the light at 4500\AA\ comes from the young stellar population and only 40\% comes from the elliptical galaxy. However, at longer wavelengths the light from the old stellar population dominates, whilst at 3400\AA\ the young population emits $\sim90$\% of the galaxy's luminosity.

The H$\alpha$ luminosity resulting from stellar photoionization is calculated from eq.\,1 in \citet{Allen95}, using the \citet{Panagia1973} ionizing fluxes and assuming unity covering fraction.  The H$\alpha$ luminosity expected for the 2- and 3-component models is 6$\times10^{40}$ and 3.4$\times10^{43}$\,erg\,s$^{-1}$, respectively. 

The total extinction-corrected H$\alpha$ luminosity of the BCG nebula is 3.4$\times10^{43}$\,erg\,s$^{-1}$, therefore between 0.2-100\% of the H$\alpha$ emitted by the galaxy is ionized by the young stellar population. 

The 3-component fit gives an upper limit to the amount of ionizing photons emitted by the young stellar population because it is forced to contain the largest possible number of O5V stars and a covering fraction of unity is asumed. Therefore the H$\alpha$ luminosity from this model is also an upper limit. So whilst it is possible that all the observed H$\alpha$ results from stellar photoionization, there is ample room for additional sources to contribute to the ionization of the nebula.

\subsubsection{Distribution of the stellar populations and nebula} 
Fig.\,\ref{fig:ifu_continuum}a displays an unsharp-mask image of the BCG and B2, created by subtracting a smoothed image from the VIMOS $R-$band image.  The galaxy is not smooth, but contains a bright filament that traverses the prominent nuclear region and extends NE to SW across the galaxy.

The IFU data is used to visualize the same R-band continuum without any emission line contamination from the H$\beta$, [O{\sc iii}] and [N{\sc i}] emission lines (Fig.\,\ref{fig:ifu_continuum}b). The centre of the BCG is estimated from this image and marked by a cross. The red continuum is centrally concentrated in an elliptical shape  and there are no bright filaments extending southwest and northeast. Therefore the large filament must be due to the line emission that falls within the R-band passband. The nearby galaxy B2 is also visible at position ($-20,0$). 

Fig.\,\ref{fig:ifu_continuum}c shows the BCG as seen in the shortest wavelength emission measured in the IFU spectra (3885--4030\AA), which in the rest-frame of the cluster is $\sim$3190--3310\AA, falling approximately in the U-band. This wavelength range does not include any bright emission lines so the light is emitted from the young stellar population. Both NE and SW filaments are prominent in the blue continuum image, implying they are the locations of the recent star formation. 

Fig.\,\ref{fig:ifu_continuum}d maps (in \oii) the emission line nebula that surrounds the BCG and shows that both the NE and SW filaments are clearly visible. The brightest region is the galaxy nucleus, and the general shape of the nebula follows the blue continuum. 

We highlight the differences between the blue and red continuum emission with Figs.\ref{fig:ifu_continuum}e and f which display the rest-frame $B-R$ continuum, and the strength of the 4000\AA\ break, D4000. Fig.\,\ref{fig:ifu_continuum}e shows that the BCG generally has blue colours, but the nucleus and SW filament are clearly bluer than the rest of the galaxy. We note that the colour variations cannot all result from the applied extinction correction. The highest H$\alpha$/H$\beta$ ratios were observed in the nuclear region (see the $E$($B-V$) map in Fig.\,\ref{fig:ebv}), which translates into a large extinction correction. Therefore the enhanced blue colours from the nucleus may result from an excessive extinction correction, however, the dust  is patchy and did not extend along the SW filament.

The variation in D4000 across the BCG is shown in Fig.\,\ref{fig:ifu_continuum}f. D4000 is greater than 1.5  in the eastern region of the galaxy, although the region with ${\rm D4000}>1.7$ at (-20,0) is the continuum from the nearby galaxy B2. The central and southwest region of the galaxy has a shallow 4000\A\ break with ${\rm D4000}<1.3$, much lower than observed in elliptical galaxies. D4000 is low in the nuclear region ($\sim1.1$) and decreases smoothly down the SW filament of young stars to the western tip of the emission line nebula, 42\,kpc away. D4000 is low in these regions, independent of the extinction correction, supporting the above finding that the central and SW parts of the galaxy host the young stars.

In summary, the old stellar population of the BCG lies within an elliptical shape. Extending beyond this smooth profile are two filaments, extending both NE and SW, which contain young, blue stars. These filaments are visible in the  {\emph{U}}-band continuum and in line emission.  There is a smooth decrease in the strength of D4000 for 42\,kpc along the SW filament implying progressively younger populations along the filament, or a larger fraction of younger stars compared to the underlying older stellar population.


\subsection{The ionized nebula surrounding the BCG}

\begin{figure*}
 \begin{center}
\centering
 \includegraphics[width=1\columnwidth]{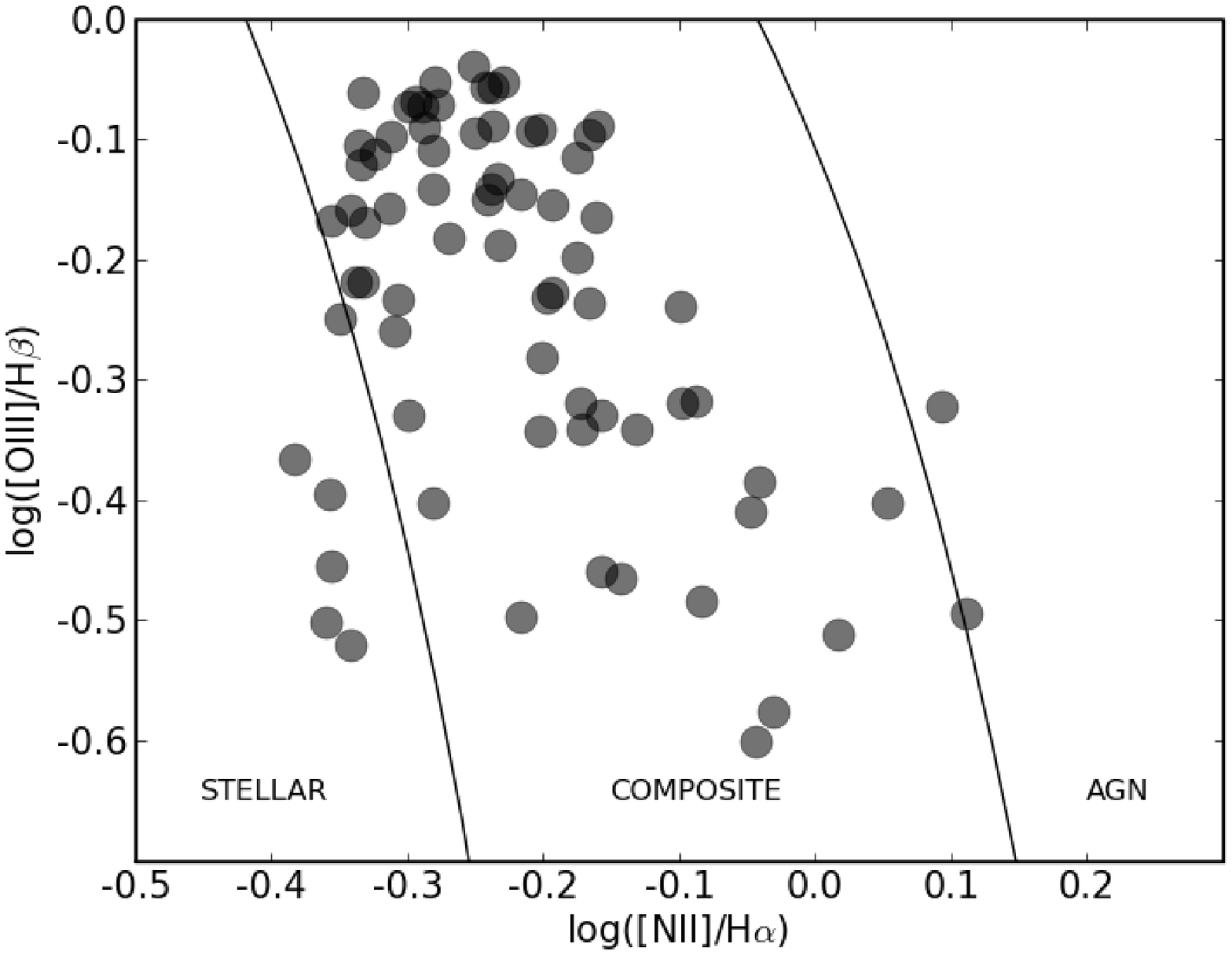}
 \includegraphics[width=1\columnwidth]{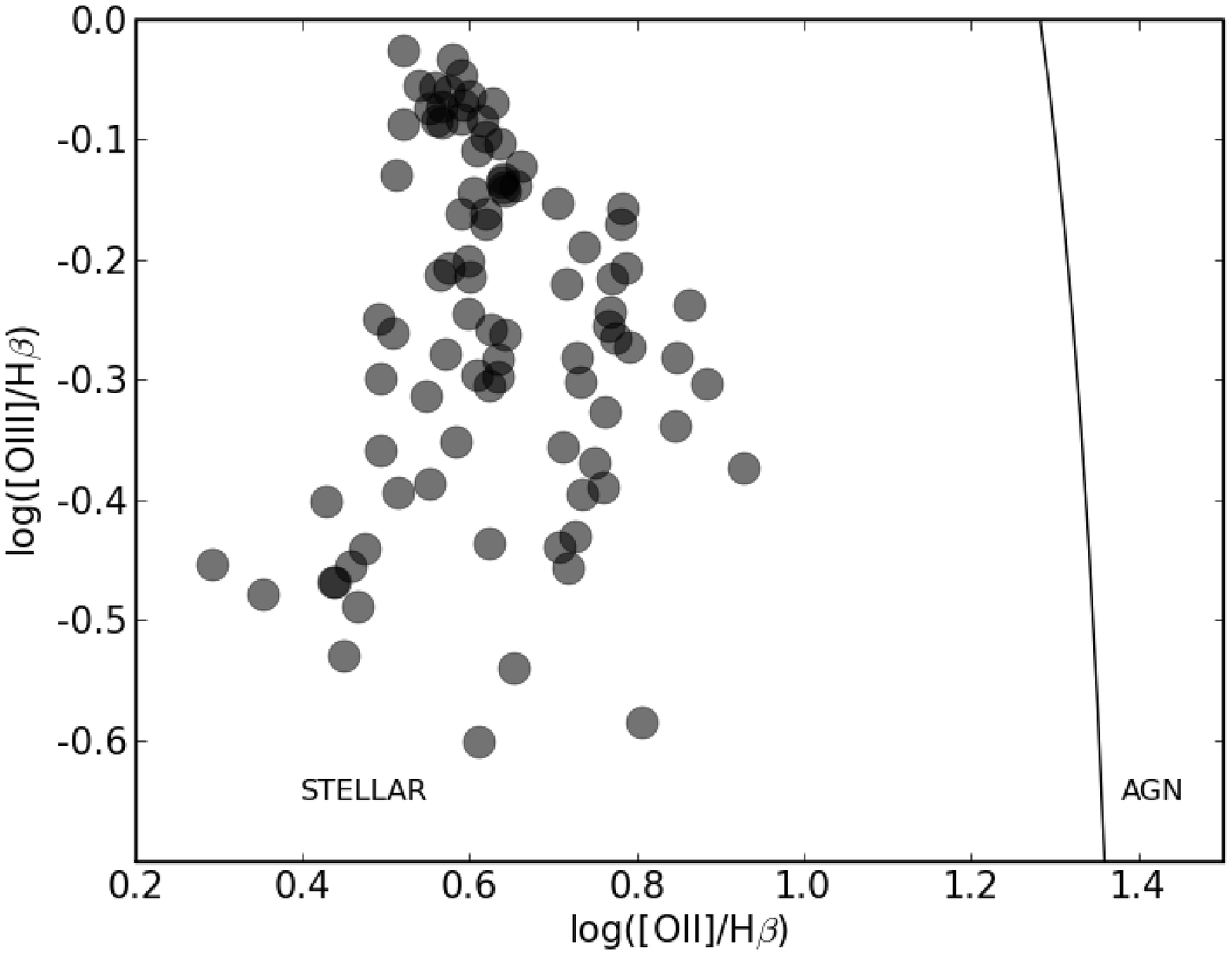}
 \caption{Diagnostic emission-line diagrams showing [N{\sc ii}] $\lambda$6584/H$\alpha$ \emph{vs.} [O{\sc iii}] $\lambda$5007/H$\beta$ and [O{\sc ii}] $\lambda$3727/H$\beta$ \emph{vs.} [O{\sc iii}] $\lambda$5007/H$\beta$ in logarithmic space. Separation lines used to differentiate between AGN, composite and stellar ionization are from \citet{Kewley2006} and \citet{Lamareille2004}, respectively. 
  \label{fig:BPT}}
\end{center}
\end{figure*}

\label{ssec:ionized_nebula}


\subsubsection{Luminosity}
\label{sssec:luminosity}
The flux in the [O{\sc ii}] line from the entire ionized nebula is $9.8\times 10^{-14}\,{\rm erg\, s^{-1}\, cm^{-2}}$ so it is more luminous than any nebula in the \citet{Crawford99} sample of BCGs (however it should be noted that the \citealt{Crawford99} fluxes luminosities are measured within slits and may be incomplete). Correcting the spectrum for reddening and extinction boosts the [O{\sc ii}] luminosity to $4.7\times10^{43}\,{\rm erg\, s^{-1}}$ making it the most luminous nebula around a BCG yet observed.

A Galaxy Evolution Explorer ({\it GALEX}) GR4 observation gives $M_{NUV}=18.78$ and $M_{FUV}=18.36$ (AB) for the BCG. In both NUV and FUV images, the emission is spatially extended along the direction of the filament. The observed FUV flux is brighter than expected given the NUV flux as dust extinction generally produces NUV$>$FUV. Since the Ly$\alpha$ emission line is shifted into the FUV passband, it is likely that the observed FUV emission is enhanced because of contamination from Ly$\alpha$. 

The UV slope $\beta$, defined as $f_{\lambda}=\lambda^{\beta}$, ranges between $-2.1$ and $-2.6$ for a galaxy which has been continuously forming stars for more than a Gyr \citep{Leitherer1999}. Almost all other stellar population models result in a higher $\beta$, therefore we can assume that $\beta>-2.6$ in this BCG. Thus from the observed NUV flux, we derive an excess flux of $1.3\times10^{-13}$\ergpscmps within the observed FUV passband. Correcting for Galactic and intrinsic extinction, we derive a lower limit of the Ly$\alpha$ emission of $3\times10^{-12}$\ergpscmps, translating into a luminosity of $3.8\times10^{44}\,\,{\rm erg\,s^{-1}}$. This is approximately a factor of 10 greater than the H$\alpha$ flux, and results in a Ly$\alpha$/H$\alpha$ ratio similar to the theoretical case A prediction.


\subsubsection{Mass and volume-filling factor}
\label{ssec:mass}

The X-ray derived temperature and central electron density of the ICM are $kT\approx 5.6\,\,{\rm keV}$ and $n_{\rm e0}=0.13\,\,{\rm cm^{-3}}$ \citep[see Table \ref{tab:rgs} and][]{Boehringer1504}. In a pure hydrogen, completely ionized gas approximation, the ICM pressure is therefore $\sim 2\times 10^{-9}\,\,{\rm dyn\,cm^{-2}}$. Thus, assuming hydrostatic equilibrium between the $10,000\,\,{\rm K}$ nebula and the ICM, the electron density inside the nebula is $n_{\rm e} \sim 850$ ${\rm cm^{-3}}$.

The mass of the ionized gas is given by
\begin{eqnarray}
 M = \frac{L_{{\rm H}\alpha}\,m_{\rm p}}{n_{\rm e}\,\alpha_{{\rm H}\alpha}^{\rm eff}\,h\nu_{{\rm H}\alpha}},
\end{eqnarray}
where $n_{\rm e}$ is the electron density, $\alpha_{{\rm H}\alpha}^{\rm eff}$ is the effective recombination coefficient and $h\nu_{{\rm H}\alpha}$ is the energy of a photon at the frequency of H$\alpha$ \citep{Osterbrock06}. For $n_{\rm e} \sim 850\,{\rm cm^{-3}}$, the equation above yields a mass of approximately $2.5\times 10^{7}$ \Msun.

Assuming this mass is distributed inside a sphere, the radius of the sphere is $\sim 50 \,\,{\rm pc}$. On the other hand, the emission in this limited region is spread over a circle of radius $\sim 20\,{\rm kpc}$. Assuming again a spherical distribution, the gas fills a volume of $\sim 3.4\times 10^{13}\,{\rm pc^3}$. Hence, the filling factor is approximately $1.5\times 10^{-8}$, a value comparable to those found by \citet{Hatch07} for four BCGs at different redshifts. This low filling fraction suggests either a clumpy or filamentary distribution of the optically-emitting gas.


\subsubsection{Source of ionization}
\label{sssec:ion} 

Shocks, conduction from the X-ray emitting ICM and UV ionization from hot young stars or the AGN may all play a role in heating  and ionizing the gas. But the line emission may also be heated by secondary electrons from more energetic particles or cosmic rays, proposed by \citet{Ferland2009} and shown to reproduce the observed line ratios of BCG nebulae. Here we study the emission line ratios in order to determine the dominant ionization and excitation source of the gas.

\citet*{BPT81} designed a method to discriminate between H{\sc ii} regions, power-law ionization and shock heating. The BPT diagrams \citep{BPT81} we use are displayed in Fig. \ref{fig:BPT}. The nebula exhibits a variety of ratios depending on the location, but no region is dominated by AGN heating, and few regions are dominated by stellar UV ionization. Rather, these diagnostic diagrams show that the gas is ionized by multiple sources. The spatial distribution of the N[{\sc ii}]/H$\alpha$, O[{\sc ii}]/H$\beta$ and O[{\sc iii}]/H$\beta$ are shown in Fig. \ref{fig:ifu_lines}. We use these figures to explore whether different ionization sources are located in certain regions.

The [O{\sc iii}]/H$\beta$ flux ratio, a good indicator of the ionization parameter, decreases radially towards the outskirts of the nebula, but is slightly offset from the centre of the galaxy. The [O{\sc iii}] may be ionized by the AGN in the nucleus. This distribution is not shared by any other line ratio, but is similar to the distribution of the [O{\sc ii}] flux.  

The [O{\sc ii}]/H$\beta$ ratio is relatively constant around the nucleus and along the protruding filament. South of this filament, the ratio suddenly increases to $\geq 0.7$, while East from the nucleus and towards B2 it decreases below 0.5. These low [O{\sc ii}]/H$\beta$ ratios are consistent with stellar UV photoionization. If an AGN dominates the ionization in the central part of the BCG, not only would we expect higher \oii/\hb\ ratios at that position, but also a gradual decrease of this line ratio towards the outskirts, where photoionization from massive stars should surmount AGN ionization. This is not observed, so we rule out this hypothesis.

\begin{figure}
\begin{center}
  \includegraphics[width=1.0\columnwidth]{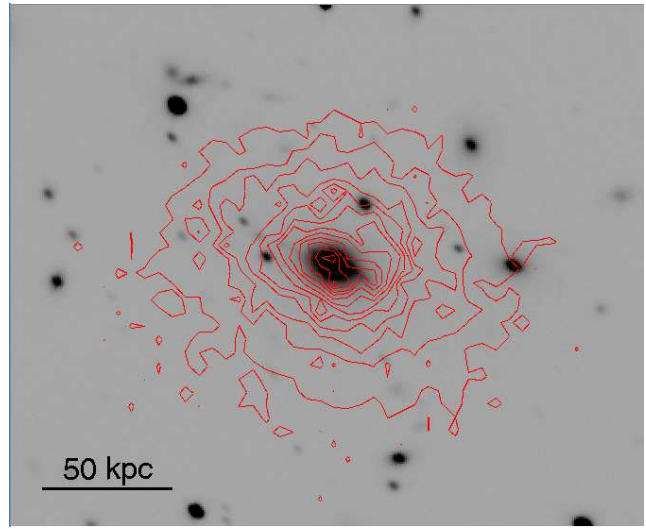}
  \caption{VIMOS R-band image with overplotted contours of the X-ray surface brightness from a \emph{Chandra} image of \citet{Boehringer1504}.\label{fig:Xray_contours}}
\end{center}
\end{figure}

The [O{\sc ii}]/H$\beta$ flux  is higher towards the South, suggesting the existence of a harder, non-stellar excitation source in this region. Fig. \ref{fig:Xray_contours} displays the X-ray surface brightness contours from \citet{Boehringer1504} overplotted on top of the VIMOS image. Two X-ray maxima are evident: one overlapping the BCG center, the other coinciding with the region of high [O{\sc ii}]/H$\beta$. Thus this hard ionization source is also visible in X-rays.

East of the nucleus, the [N{\sc ii}]/H$\alpha$ ratio varies rapidly between -0.35 and 0.15 (in logarithmic space), which is too large to be due to stellar photoionization. Possible heating mechanisms include shock heating or power-law ionization. To the West, however, this flux ratio is consistently low up to a distance of $\sim 15$ kpc from the centre; then there is a slight increase again in the SW tip of the nebula. The region of low [N{\sc II}]/H$\alpha$ flux ratios coincides with that of a low D4000 and a blue B-R colour, suggesting the gas in this region is ionized by UV radiation from young stars. However the BPT diagram (Fig. \ref{fig:BPT}a) indicates that this region is heated by more than one source, so an additional source competes with the UV from young stars to ionize the gas.

\begin{figure}
\begin{center}
  \includegraphics[width=1.0\columnwidth]{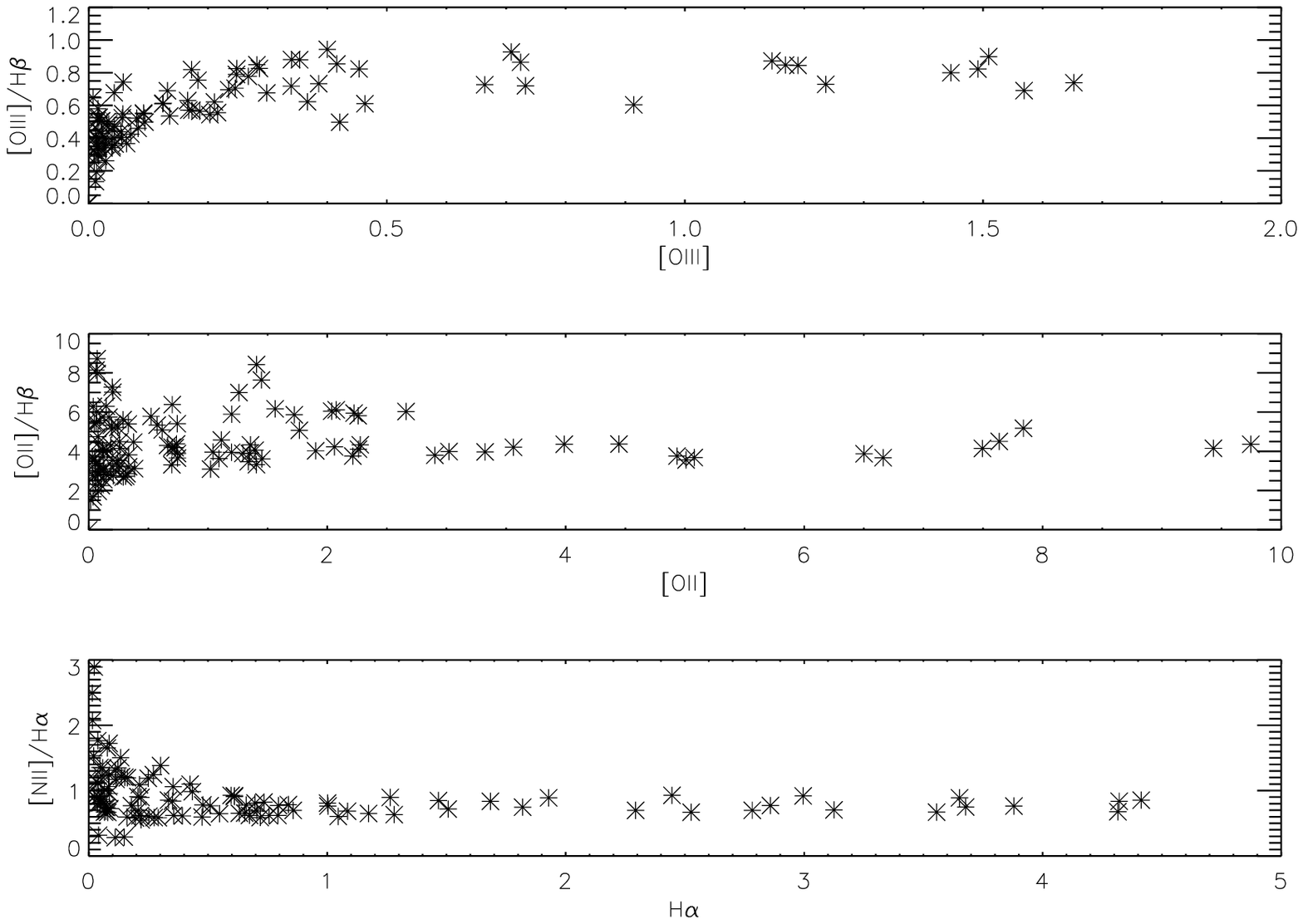}
  \caption{[O{\sc iii}] \emph{vs.} [O{\sc iii}]/H$\beta$, [O{\sc ii}] \emph{vs.} [O{\sc ii}]/H$\beta$ and H$\alpha$ \emph{vs.} [N{\sc ii}]/H$\alpha$. All values on the x-axis are fluxes in units of $10^{-14}$ \ergpscmps.\label{fig:saturated}}
\end{center}
\end{figure}

Fig. \ref{fig:saturated} shows the three line-ratios as a function of line flux. The line ratios saturate at high luminosities, whereas fainter regions are characterized by more variability.  This trend has also been observed in other luminous BCGs by \citet{Wilman2006} who postulate that the line ratios saturate when star formation is proceeding at such a high rate that stellar UV dominates the photoionization of the gas.

In conclusion stellar UV is a plausible ionizing source, but it cannot act alone: harder ionizing sources seem to be also present. The stellar UV ionization is concentrated along the SW filament, whilst the ionizing sources of the nuclear and other regions remain unknown. The ionizing source of the nuclear region is of particular importance as this region is brightest in both the U-band and line emission. Our data point towards ionization by multiple sources, including stellar UV.


\subsubsection{Gas kinematics}
\label{sssec:kin}

\begin{figure*}
 \begin{center}
\centering
\includegraphics[width=0.79\columnwidth]{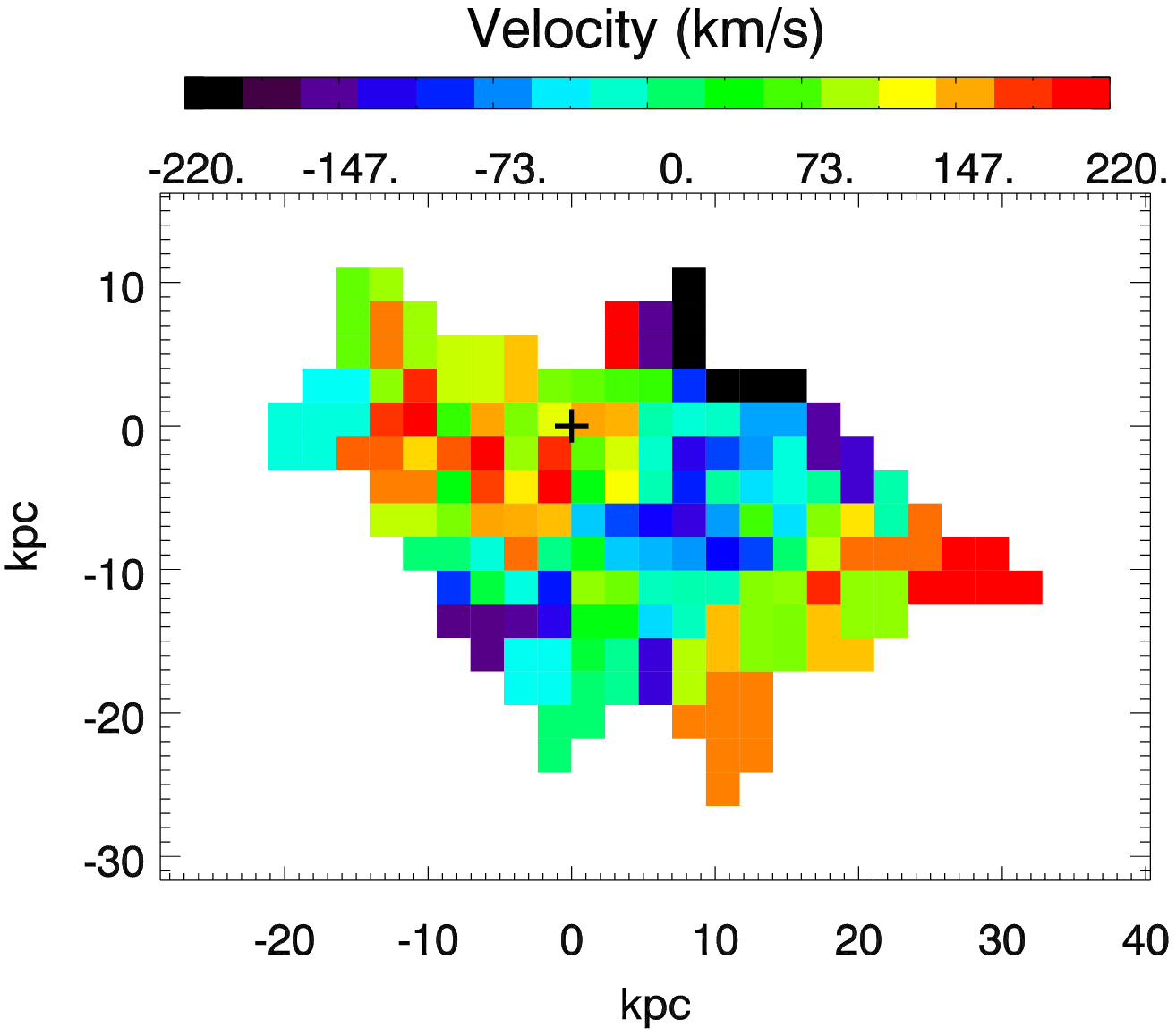}
  \includegraphics[width=0.79\columnwidth]{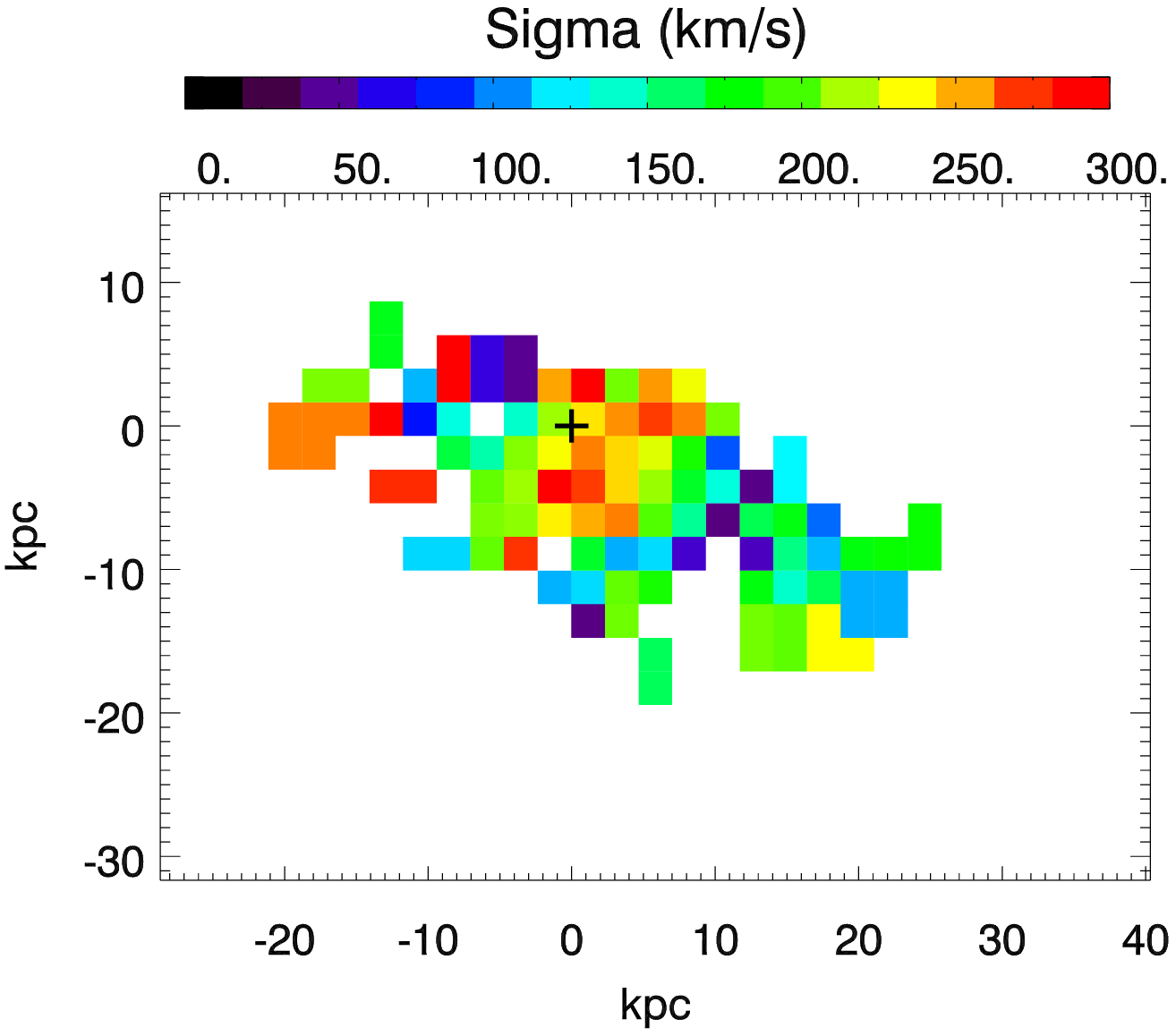}
 \caption{Kinematic properties of RXJ1504, from left to right: radial velocity derived from the redshift of the [O{\sc ii}] doublet and velocity dispersion $\sigma$, both in units of \kmps. The zero velocity point is defined as the redshift of the [OII] emission line measured in the combined spectrum shown in fig\,\ref{fig:spectrum}. 
   \label{fig:ifu_kinematics}}
 \end{center}
\end{figure*}

Fig. \ref{fig:ifu_kinematics}a presents the line-of-sight kinematics of the nebula, as derived from the relativistic Doppler shifts of the [O{\sc ii}] emission lines. All velocities were computed with respect to the median redshift of the [O{\sc ii}] $\lambda 3727$ emission ($z=0.2173$). 

The galaxy is broken up into distinct regions. The galaxy B2 has a gas velocity of $-50$\kmps, and is clearly distinct from the red-shifted gas that lies to the East of the BCG and in the nuclear region, which has a velocity between 0 and 220\kmps.  To the West of the nuclear region the gas is blue-shifted up to $-220$\kmps, although at the western-most tip of the nebula the gas velocity again changes direction and is red-shifted by 200\kmps. The central part of the SW filament is blue-shifted, whilst the SW tip is redshifted, thus the filament must either be stretching or collapsing on itself.  Similar kinematics can be seen in NGC\,1275 in the Perseus cluster \citep{Hatch06}, and in the models of \citet{Pope2008}. The length of the SW filament is 42 kpc, implying a dynamical age of $\sim10^8$\,yr, so the line-emitting gas must be long-lived.

The dispersion map of the gas is displayed in Fig. \ref{fig:ifu_kinematics}b.  All line-widths are $\sigma$-widths, and not FWHM (full width at half maximum).  The $\sigma$-widths of the [O{\sc iii}] and H$\beta$ lines used for creating the dispersion map of the gas were corrected for instrumental broadening of 375\kmps\ as measured from the nearby 5755\AA\ skyline. White bins in this figure are regions in which the spectral resolution is too low to determine a dispersion. A peak in the dispersion is seen slightly West of the nuclear region. The dispersion here is 250-300\kmps\ whilst the median velocity dispersion is $\sim$200\kmps. The SW filament has the lowest velocity dispersion. These strong internal motions may result from the superposition of distinct components along the line of sight. 


\subsection{Star formation rate of the BCG}
\label{sec:sfr}

The FUV light from a galaxy is dominated by emission from young O and B type stars so we can use the rest-frame FUV emission to estimate the star formation rate. This estimate of the SFR does not assume that the line emission has been ionized by the young stellar population. Since the NUV and FUV emission is spatially extended, approximately in the same direction as the filament, it is likely to be emitted from a spatially extended source, such as stars, rather than the central active nucleus.

Assuming all of the 1800\AA\ flux is emitted from a population of hot young stars, we use the observed NUV emission (equivalent to the rest-frame 1870\AA) to obtain a star formation rate. The NUV flux of $1.1\times10^{-27}\,{\rm erg\, s^{-1}\, cm^{-2}\, Hz^{-1}}$ is corrected for Galactic extinction of $E$($B-V$)$=0.108$ and internal extinction of $E$($B-V$)$=0.211$. We use the relation between star formation rate and rest-frame FUV flux (observed NUV flux) derived by \citet{Salim2007} assuming a Salpeter mass function.
 
\begin{equation}
 SFR_{\rm FUV} = 1.08\times 10^{-28}\, L_{\rm FUV} \,\,\,\, {\rm M}_{\sun} {\rm \, yr^{-1}}.
\end{equation}

This equation is calibrated for FUV emission centered on 1528\AA, whilst the rest-frame wavelength of the light observed through the NUV filter is 1866\AA. However \citet{Salim2007} show that this will have a negligible effect on the conversion between FUV luminosity and star formation rate. Star formation rates derived from the observed NUV flux are given in Table \ref{tab:sfr}. \citet{Kennicutt98} derive a relation that results in 30\% greater star formation rate for a given NUV flux.

The H$\alpha$ and [O{\sc ii}] luminosity may also be used to estimate the star formation rate in the BCG under the assumption that the gas is ionized by UV photons from hot young stars. Table\,\ref{tab:sfr} lists the total flux, extinction corrected fluxes and luminosities of both [O{\sc ii}] and H$\alpha$, together with the derived star formation rates.
Star formation rates (SFR) are computed from the [O{\sc ii}] and H$\alpha$ luminosites according to the empirical relation of \citet*{Kewley04},
\begin{equation}
 SFR^{\oii}_{\rm Kw04} = 6.58\times 10^{-42}\, L_{\rm \oii} \,\,\,\, {\rm M}_{\sun} {\rm \, yr^{-1}},
\end{equation}
and the theoretical equation of \citet{Kennicutt98},
\begin{equation}
 SFR^{{\rm H}\alpha}_{\rm Kn98} = 7.8\times 10^{-42}\, L_{\rm H\alpha} \,\,\,\, {\rm M}_{\sun} {\rm \, yr^{-1}},
 \end{equation}
respectively, where $L_{\rm \oii}$ and $L_{{\rm H}\alpha}$ are expressed in units of ${\rm erg\, s^{-1}}$. 

\begin{table}
  \begin{tabular}{lcccccc}
  \hline
Line & $\mathcal{F}$ & L&SFR &Ext.-cor.&Ext.-cor. & Ext.-cor.\\
&  && & $\mathcal{F}$ & L &SFR \\ \hline
 [O{\sc ii}] & 9.8 & 1.3 & 83 & 36.9 & 4.7 & 314 \\
 H$\alpha$&14& 1.8 & 141 & 26.2 & 3.4 & 262 \\ 
   FUV&1.10&--&15.1&11.8&--&136\\ \hline
\end{tabular}
  \caption{Fluxes ($\mathcal{F}$) are stated in units of $10^{-14}\,{\rm erg\, s^{-1}\, cm^{-2}}$, FUV flux density is in units of $10^{-27}\,{\rm erg\, s^{-1}\, cm^{-2}\, Hz^{-1}}$, luminosties (L) are stated in units of $10^{43}\,{\rm erg\, s^{-1}}$ and star formation rates (SFR) are given in units of \Msunpyr. \label{tab:sfr}}
\end{table}

The FUV-derived star formation rates (calculated from the observed NUV emission) are 2--3 times lower that the rates derived from the H$\alpha$ and [O{\sc ii}] emission lines. In section \ref{sec:stellarpop} we show the number of O and B stars present in the BCG is barely sufficient to power the H$\alpha$ emission. The discrepancy between the star formation rates derived from FUV and emission line luminosity reinforces this conclusion  and implies that the [O{\sc ii}], H$\alpha$ and other emission lines are heated by an additional source. Star formation rates of BCGs derived from line emission must only be considered upper limits of the true star formation rates.


\section{Discussion}

\label{sec:disc}

\subsection{Brightest cluster galaxy growth}
We present a number of observations which imply the BCG is rapidly forming stars. These include extended FUV emission, blue galaxy colours atypical of elliptical galaxies, shallow D4000, and some line ratios indicating stellar UV as a possible ionization source.
The IFU continuum and line ratio maps indicate this star formation occurs mainly in the galaxy core and in a 42\,kpc filament stretching from the core to the SW.

The rest-frame FUV emission is likely to be the most reliable estimator of the star formation rate, thus setting it at $\sim$136\Msunpyr. This approximately agrees with the mass deposition rate derived from the X-ray spectra ($\sim 80$\Msunpyr). Therefore, the current star formation may be fueled by condensing intracluster gas.

The current star formation rate is large, and together with the BCG of Abell 1835, it is the largest star formation rate of any low-redshift elliptical galaxy. However, this current rate is considerably less than the average rate at which stars must have formed over the galaxy's entire lifetime. The specific star formation rate of the BCG (the ratio of current SFR to the mass of galaxy) is $\sim2.3\times10^{-11}$\pyr\ and is therefore similar to many elliptical galaxies which are detected in the UV. Had the current rate of star formation lasted since $z=1$, the galaxy would have increased its mass by less than 15\%. If this phase of star formation should only last a few tens of Myrs, the galaxy would return to its pre-starburst colours in a short time after the end of the burst. Thus, while the ongoing star formation clearly influences the galaxy's colours and ionization state of the gas, its impact on the mass of the galaxy is not significant. 

\subsection{Heating of the nebula}

The optical emission line ratios indicate stellar photoionization dominates in the centre of the galaxy and along the star forming filaments. The strong blue continuum and shallow D4000 in these regions indicate the presence of ionizing O and B stars.

However, the BPT diagnostic diagrams show stellar photoionization cannot be the exclusive ionization source and at least one additional mechanism needs to be present. 

Both H$\alpha$ and [O{\sc ii}] line luminosities estimate higher star formation rates relative to the FUV derived rate. The number of ionizing O and B stars in the BCG, inferred from SED fitting, is barely sufficient to produce all the H$\alpha$ flux emitted. This also implies that some of the line emission is ionized by an additional source. Thus the H$\alpha$ or [O{\sc ii}] flux should only be used to calculate an upper limit of the star formation rate. We have also shown that dust extinction can be large in BCGs, hence the observed line flux will be greatly reduced, and star formation rates derived from single emission lines should be considered highly uncertain.

Line ratio maps are a good means to detect where the dominant ionization source changes. The best ratio for this work is [O{\sc ii}]/H$\beta$ because the [N{\sc ii}] and H$\alpha$ lines lie in  a wavelength region that is badly affected by fringing, so the ratio of these lines is unreliable.  The  [O{\sc ii}]/H$\beta$ map clearly marks the change in dominant ionization source in the South, which coincides with a second X-ray peak. None of the line ratio maps show a correlation with the map of recent star formation (Fig.\ref{fig:ifu_continuum}c), providing further evidence that stellar photoionization is not the only ionizing source at work. This result does not come as a surprise and it has also been observed in other BCGs \citep[e.g.][]{Sabra2000,Hatch07}. \citet{Ferland2009} have proposed an alternative particle heating method which reproduces the line ratios.

\subsection{AGN Feedback}

The mass deposition rate allowed by the X-ray spectra is less than 10 percent of the mass deposition rate of 1400-1900 \Msunpyr implied by the total X-ray luminosity emitted from the cluster core. Since we have an indication that 10 percent of the ICM is cooling and forming stars, it follows that a heating source must supply 90 percent of the luminosity emitted from within the cooling radius, which is $2.7\times 10^{45}$ \ergps. If this is supplied by AGN feedback from the central black hole, the mass accretion rate must be 0.5 \Msunpyr, assuming at least a 10 percent output efficiency.

The $K-$band magnitude of a galaxy correlates with the mass of the supermassive black hole (SMBH) at its centre \citep[e.g.][]{Marconi2003,Graham2007}:
\begin{equation}
  \log \left(M_{\rm BH}/\Msun\right)=-0.33\,\left[M_K+24\right]+8.33,
\end{equation}
where ${M_{\rm BH}}$ is the mass of the black hole and ${M_{K}}$ is the {\emph{K}}-band magnitude of the BCG. With ${M_{K}}=-27.0$ (2MASS), it follows the BCG has a SMBH of mass $\sim 2\times 10^9$ \Msun.

To estimate the power supplied by an AGN, we first consider Bondi accretion from the hot gas \citep{Bondi1952}. The Bondi accretion radius is $\sim 12$ pc. As for all SMBH except perhaps the one in M87, this is far beyond the resolution limit of our instruments and we can only speculate about the physical conditions close to the black hole. The black hole will accrete mass at a rate $\dot{M}_{\rm Bondi}=\pi\lambda c_{\rm s} \rho r_{\rm A}^2$, where $\lambda$ depends on the adiabatic index \citep[here $\gamma=5/3$, $\lambda=0.25$;][]{Bondi1952}, $c_{\rm s}$ is the sound speed, $\rho$ is the gas density and $r_{\rm A}$ is the accretion radius. For the central parameters determined from the X-ray data, namely $kT=5.6$ keV and $n_e = 0.13\,\,{\rm cm^{-3}}$, we obtain  $\dot{M}_{\rm Bondi}\approx 5\times 10^{-4}$ \Msunpyr. However, this temperature and density represent an average over a large extraction region used for the X-ray spectral analysis, therefore the true temperature and density near the black hole could be very different. If we assume that the thermal gas pressure at the centre is no less than the thermal pressure of the inner region measured spectrally to be at a temperature of 5.6 keV, then $\rho$ scales as $1/T$ (in practice the extra weight will make it go steeper than this). Therefore, $\dot{M}_{\rm Bondi} \propto 1/T^{2.5}$. So to increase $\dot{M}_{\rm Bondi}$ by $10^3$, to 0.5 \Msunpyr, requires only a drop in temperature by a factor of 16 which means to 350 eV. Many nearby cool core clusters show a cool component at $\sim 0.5$ keV \citep[e.g.,][]{Sanders2010} and we cannot rule out such a low temperature in the centre of RXCJ1504.

One could also try to infer the AGN power output over the past $10^8$ years from the radio power at 1.4 GHz using the relation of \citet{Birzan2008}:
\begin{equation}
\log P_{\rm jet} = (0.35)\,\log P_{\rm 1400} + 1.9 .
\end{equation}
In the relation above, $P_{1400}$ represents the bolometric radio luminosity for the total source at 1.4 GHz in units of $10^{24}$ ${\rm W\, Hz^{-1}}$, and $P_{\rm jet}$ is the jet/cavity power in units of $10^{42}$ ${\rm erg\,s^{-1}}$.

The BCG contains a radio source with a brightness of 62 mJy at 1.4 GHz \citep{Bauer2000}. At a luminosity distance of $1075.5$ Mpc, this implies a radio power of $8.6\times 10^{24}$ ${\rm W\,Hz^{-1}}$, thus a jet power of $1.5\times 10^{44}\,\,{\rm erg\,s^{-1}}$. The corresponding black hole accretion rate at 10 percent efficiency is 0.03 \Msunpyr and an order of magnitude smaller than the power required to heat the ICM. However, the relation found by Birzan has a scatter of 0.85 dex (factor of $\sim 7$) at 1400 MHz, so that our object lies a little over $1 \sigma$ away from the mean ($\sim 1.4 \sigma$ in log).

In conclusion, we cannot rule out that heating is supplied by a central AGN that accretes with a rate of $\sim 0.5$ \Msunpyr.

If the accretion rate has been as high as 0.5\Msunpyr since redshift 1, then the black hole has grown by an amount of $\sim 3 \times 10^9$ \Msun, which is larger than the total estimated mass of $\sim 2\times 10^9$ \Msun\ . This implies that the current accretion rate is atypically high, which suggests that the present phase in the feedback cycle with the high star-formation and accretion rate must be relatively short lived. Furthermore, BCGs may harbor supermassive black holes with masses a few times higher than those predicted by the $M_{\rm BH}-M_{\rm K}$ relation of \citet{Graham2007}, as is the case in Abell 1836 \citep{Dalla2009}.


\subsection{Turbulent energy of the ICM}
\label{ssec:nonth}

Assuming the velocity dispersions and the line-of-sight velocities of the cold gas are solely due to turbulence gives an upper limit on the turbulent velocities in the nebula.
Despite the large density contrast between the ICM and the optical line-emitting gas, the low filling factor suggests that the nebula gas may be blown about by the hot gas and thus serves as a good tracer of ICM motions \citep{Fabian2003}. The speed of sound in the hot central plasma at a temperature of 5.6 keV is roughly 1190 ${\rm km\,s^{-1}}$ and the characteristic velocity of isotropic turbulence is $v_{\rm turb}= \sqrt{2\,\left(v_\Delta^2 + v_\sigma^2\right)}$, where $v_\Delta$ is the dispersion of the velocity shear of the cold gas, measured from the $\sigma$-width of the line-of-sight velocity histogram, $v_\Delta\approx 117\,\kmps$, and $v_\sigma$ is the average velocity dispersion, $v_\sigma\approx 200\,\kmps$. $v_{\rm turb}$ provides an upper limit on the velocity dispersion of the hot gas. $v_{\rm turb}\approx 325\,\kmps$ and therefore the turbulence in the central ICM has a typical Mach number of less than 0.27. We thus obtain an upper limit on the ratio of turbulent to thermal energy in the ICM, given by $(\gamma/2)\:{\rm M}^2$ with $\gamma=5/3$ the adiabatic index of monatomic ideal gas, of roughly 6\%, in agreement with upper limits derived from other methods \citep[]{Churazov08,Werner2009}.

\section{Conclusions}
\label{sec:concl}

RXCJ1504.1-0248 is one of the most extreme cooling flows ever discovered. The observations presented here offer the first multi-wavelength view of the cluster core, examining both the hot and warm gas as well as the stars of the BCG. 

The core of RXCJ1504.1-0248 has an X-ray luminosity of at least $3\times 10^{45}$ ${\rm erg\, s^{-1}}$, approximately 10 percent of which is supplied by gas cooling and forming stars. Accretion onto a supermassive black hole of $2\times 10^9\,\, \Msun$ is the likely mechanism needed to account for most of the remaining $2.7\times 10^{45}$ ${\rm erg\,s^{-1}}$ heating. The corresponding Bondi accretion rate at 10 percent radiative efficiency is 0.5 \Msunpyr. Using the correlation between 1.4 GHz radio luminosity and jet power \citep{Birzan2008}, we infer a jet power of $1.5\times 10^{44}$ ${\rm erg\, s^{-1}}$, still insufficient to heat the ICM.

The intracluster medium mass deposition rate is in agreement with the star formation rate of the BCG derived through FUV observations ($\sim$140\Msunpyr). The IFU images show that most of the star formation occurs in the galaxy core and in a 42\,kpc-long filament that stretches out from the core to the SW. These regions have blue continuum colours, and shallow 4000\AA\ breaks consistent with recent star formation. Additionally, the [O{\sc ii}]/H$\beta$ line ratio indicates stellar photoionization is the dominant source of ionization along the filament. The X-ray emission also extends along the filament indicating a direct connection between the intracluster medium and the recent star formation.

The line-emitting nebula that surrounds the BCG is the most luminous observed to date, with extinction-corrected [O{\sc ii}]  and H$\alpha$ luminosities of $4.7\times 10^{43}\,\,{\rm erg\,s^{-1}}$ and $3.4\times 10^{43}\,\,{\rm erg\,s^{-1}}$ respectively. The number of ionizing O and B type stars in the BCG is barely sufficient to power this bright nebula, and the BPT diagrams and line ratio maps show that an additional heating mechanism is required. Because this heating source produces H$\alpha$ emission, BCG star formation rates that are derived from H$\alpha$ luminosities should be considered upper limits. 

The nebula kinematics are ordered and the nebula comprises of a number of kinematically distinct regions.  The velocities are generally low ($<250\,\,{\rm km\,s^{-1}}$), and there is no evidence for rotation or free-falling gas infall. The 42\kpc\ filament has a velocity shear of  $\sim$400\kmps\ resulting in a dynamical time of  $10^{8}$\,years, thus the nebula must be long-lived. The average velocity dispersion of the nebula is $v_{\sigma} \sim 200$ ${\rm km\,s^{-1}}$ and the dispersion of the velocity shear of the cold gas is $v_{\Delta}\sim 117\,\kmps$. These set an upper limit of $\sim 325\,\kmps$ on the mean turbulent velocity in the ICM, meaning that the ratio of turbulent to thermal energy of the intracluster medium is limited to less than 6 percent.

Our results suggest mass transfers from the intracluster medium to the brightest cluster galaxy at a rate of approximately $10^{2}$\Msunpyr. The cooling of the intracluster medium results in a gas reservoir in the brightest cluster galaxy. The rate at which gas condenses into this reservoir is similar to the rate at which this reservoir forms stars, so the reservoir should be approximately stationary. Our observations cannot reveal the coolest gas components of this reservoir, but we find that the $10^{4}$\,K gas is partly heated by a source other than the forming stars. 

\section*{Acknowledgments}

We thank the referee for many helpful comments. The VLT-VIMOS Integral Field Unit data presented in this paper were reduced using the VIMOS Interactive Pipeline and Graphical Interface (VIPGI) designed by the VIRMOS Consortium. We thank Bianca Garilli and Luigi Paioro for their assistance in using VIPGI. This work is partly based on observations obtained with XMM-Newton, an ESA science mission with instruments and contributions directly funded by ESA member states and the USA (NASA). NAH acknowledges funding from the Royal Netherlands Academy of Arts and Sciences. HB acknowledges support through the Cluster of Excellence ``Origin and Structure of the Universe'', funded by the Excellence Initiative of the German Federal Government as EXC project 153. AS and MB acknowledge support by the DfG Schwerpunkt programme SP1177. NW is supported by the National Aeronautics and Space Administration through Einstein Postdoctoral Fellowship Award Number PF8-90056 issued by the Chandra X-ray Observatory Center, which is operated by the Smithsonian Astrophysical Observatory for and on behalf of the National Aeronautics and Space Administration under contract NAS8-03060.

\bibliographystyle{mn2e}
\bibliography{bibliography}

\label{lastpage}

\end{document}